\documentclass[sigconf,table,xcdraw,screen]{acmart}
\usepackage[ruled,vlined]{algorithm2e}
\usepackage{graphicx}
\usepackage{paralist}
\usepackage{tabularx}
\usepackage{balance}
\usepackage{multirow}
\usepackage{multicol}
\usepackage{pbox}
\usepackage{mathtools}
\usepackage[TABBOTCAP]{subfigure}
\usepackage{algorithmic}
\usepackage{paralist}
\usepackage{url}
\usepackage{balance}
\usepackage{multirow}
\usepackage{multicol}
\usepackage{makecell}
\usepackage{amsmath}
\usepackage{anyfontsize}
\usepackage{setspace}
\usepackage{verbatim}
\usepackage{float}
\usepackage[normalem]{ulem}
\usepackage{xspace}
\usepackage{amssymb}
\usepackage{ifthen}
\usepackage{pifont}
\usepackage{tikz}

\usepackage{hyperref}
\usepackage{amsmath, xfrac}
\usepackage{caption}
\usepackage{CJK}
\usepackage{cellspace}
\newcolumntype{M}{>{\centering\arraybackslash}m{\dimexpr.10\linewidth+8\tabcolsep}}

\renewenvironment{itemize}[1]{\begin{compactitem}#1}{\end{compactitem}}
\renewenvironment{enumerate}[1]{\begin{compactenum}#1}{\end{compactenum}}

\usepackage{enumitem}
\setlist[itemize]{leftmargin=5mm}

{

\newcommand{\ie}{\textit{i.e.},\xspace}
\newcommand{\eg}{\textit{e.g.},\xspace}
\newcommand{\etc}{\textit{etc.}\xspace}
\newcommand{\etal}{\textit{et al.}\xspace}
\newcommand{\aka}{\textit{a.k.a.}\xspace}

\newcommand{\figref}[1]{Figure~\ref{#1}\xspace}

\newcommand{\tabref}[1]{Table~\ref{#1}\xspace}

\newcommand{\crashscope}{{\scshape{CrashScope}}\xspace}
\newcommand{\approach}{{\scshape{Euler}}\xspace}
\newcommand{\approachtwo}{{\textsc{Euler}}\xspace}
\newcommand{\euler}{\approach}
\newcommand{\approachlong}{{\scshape{ass\textbf{E}ssing the 
q\textbf{U}a\textbf{L}ity of the steps to r\textbf{E}produce in bug 
\textbf{R}eports}}\xspace}
  		
\newcolumntype{L}[1]{>{\raggedright\let\newline\\\arraybackslash\hspace{0pt}}p{
#1}}
\newcolumntype{C}[1]{>{\centering\let\newline\\\arraybackslash}p{#1}}
\newcolumntype{R}[1]{>{\raggedleft\let\newline\\\arraybackslash\hspace{0pt}}p{
#1}}
	
\newcommand{\emphquote}[1]{{\emph{`#1'}}\xspace}
\newcommand{\emphdblquote}[1]{{\emph{``#1''}}\xspace}
\newcommand{\emphbrack}[1]{\emph{[#1]}\xspace}

\newcommand{\act}{\emphbrack{action}}
\newcommand{\obj}{\emphbrack{object}}
\newcommand{\prep}{\emphbrack{preposition}}
\newcommand{\objtwo}{\emphbrack{object2}}

\copyrightyear{2019} 
\acmYear{2019} 
\setcopyright{acmcopyright}
\acmConference[ESEC/FSE '19]{Proceedings of the 27th ACM Joint European 
Software Engineering Conference and Symposium on the Foundations of Software 
Engineering}{August 26--30, 2019}{Tallinn, Estonia}
\acmBooktitle{Proceedings of the 27th ACM Joint European Software Engineering 
Conference and Symposium on the Foundations of Software Engineering (ESEC/FSE 
'19), August 26--30, 2019, Tallinn, Estonia}
\acmPrice{15.00}
\acmDOI{10.1145/3338906.3338947}
\acmISBN{978-1-4503-5572-8/19/08}

\begin{document}

\title{Assessing the Quality of the Steps to Reproduce in Bug Reports}

\author{Oscar Chaparro$^1$, Carlos Bernal-C\'ardenas$^1$, Jing Lu$^2$, Kevin 
Moran$^1$, 
Andrian Marcus$^2$, Massimiliano Di Penta$^3$, Denys Poshyvanyk$^1$, Vincent  
Ng$^2$}
\affiliation{%
\institution{
$^1$College of William and Mary, USA -- $^2$The University of Texas at Dallas, 
USA -- $^3$University of Sannio, Italy
}
}

\renewcommand{\authors}{Oscar Chaparro, Carlos Bernal-C\'ardenas, Jing Lu, 
Kevin Moran, Andrian Marcus, Massimiliano Di Penta, Denys Poshyvanyk, and 
Vincent Ng}

\renewcommand{\shortauthors}{O.~Chaparro, C. Bernal-C\'ardenas, J.~Lu, 
K. Moran,\\ A. Marcus, M. Di Penta, D. Poshyvanyk, and 
V. Ng}

\begin{abstract}
A major problem with user-written bug reports, indicated by developers and 
documented by researchers, is the (lack of high) quality of the reported 
\textit{steps to reproduce} the bugs. 
Low-quality steps to reproduce lead to excessive manual effort spent on bug triage and resolution. 
This paper proposes \approach, an approach that automatically identifies 
and assesses the quality of the steps to reproduce in a bug report, 
providing feedback to the reporters, which they can use to improve the bug report.
The feedback provided by \approach was assessed by external evaluators and the 
results indicate that \approach correctly identified 98\% of the existing steps 
to reproduce and 58\% of the missing ones, while 73\% of its quality 
annotations~are~correct.
\end{abstract}

\ccsdesc[500]{Software and its engineering~Maintaining software}

\keywords{Bug Report Quality, Textual Analysis, Dynamic Software Analysis} 

\maketitle


\section{Introduction}
\label{sec:intro}

When software does not behave as expected, users and/or developers report the problems using issue trackers \cite{Zimmermann2009}. Specifically, problems are frequently reported as \textit{bug reports}, \ie documents that describe software bugs 
and are expected to contain the information needed by the developers to triage 
and fix the bugs in the software. 

Along with the observed and expected behavior, bug reports often contain the 
steps to reproduce 
 (S2Rs) the bug. The S2Rs are essential in helping developers to 
replicate and correct the bugs~\cite{Laukkanen2011, Zimmermann2010}. 
Unfortunately, in many 
cases, the S2Rs are unclear, incomplete, and/or ambiguous.  So 
much so that 
developers are often unable 
to replicate the problems, let alone fix the bugs in the software 
\cite{ErfaniJoorabchi2014,Zimmermann2012,Breu2010,Guo2010,Zimmermann2010,GitHub2016,Fazzini2018,Karagoez2017}.
Recently, developers from more than 1.3k open-source projects wrote a 
letter to GitHub expressing their frustration that the S2Rs
are often missing in bug reports 
\cite{GitHub2016}, and asking for a solution that would make reporters include 
them in the reports. In addition, prior research found that low-quality 
S2Rs lead to non-reproducible bugs~\cite{ErfaniJoorabchi2014}, 
unfixed bugs~\cite{Zimmermann2012}, and excessive 
manual effort spent on bug triage and 
resolution~\cite{ErfaniJoorabchi2014,Breu2010,Guo2010,Zimmermann2010}. 
Low-quality 
S2Rs are also the main problem with automated approaches attempting to generate test cases from bug reports \cite{Karagoez2017,Fazzini2018}.  
For example, Fazzini \etal \cite{Fazzini2018} report that a S2R may refer to an interaction outside the system or it may be ambiguous and correspond to multiple interactions.
 Similar problems were encountered by Karag{\"o}z \etal~\cite{Karagoez2017}, 
 who even proposed the adoption of a semi-formal 
 format to express S2Rs,  attempting to alleviate such issues.

Ideally, low-quality S2Rs in bug reports should be identified at reporting 
time, such that reporters would have a chance to correct them. With that in 
mind, we propose \approach,  an approach that 
automatically analyzes the textual description of a bug 
report, assesses the quality of the S2Rs, and provides 
actionable feedback to reporters about: ambiguous steps, steps described with 
unexpected vocabulary, and steps missing in the report. In this paper, we 
present the approach and evaluate an implementation geared towards 
S2Rs corresponding to GUI-level interactions in Android applications. \approach 
can be adapted to support any GUI-based system.

\approach leverages neural sequence labeling 
\cite{Huang2016LSTM,Lample2016neural} in combination with 
discourse patterns \cite{Chaparro2017-2} and dependency parsing 
~\cite{Manning2014} to identify 
S2R sentences and individual S2Rs.
Next, it matches the S2Rs to program states and 
GUI-level application interactions, represented in a graph-based execution 
model. A successful match indicates that the S2R precisely corresponds to an 
app interaction (\ie it is of high-quality). Conversely, a low-quality S2R may 
match to multiple 
screen components or app events, may not match any application state or 
interaction, or it may require the execution of 
additional steps. \approach assigns to each S2R quality annotations that 
provide specific feedback to the reporter about problems with the S2Rs.  

We asked external evaluators to assess the accuracy
and completeness of the quality reports produced by \approach 
for 24 bug reports of six Android applications.  The results indicate that: 
\approach correctly identifies 98\% of the S2Rs; 73\% of \approach's quality 
annotations are correct; and \approach successfully identifies 58\% of the 
missing S2Rs. The evaluators also provided feedback on the perceived usefulness 
of the information included in the quality reports, on the additional 
information that should be in them, as well as on usability. The 
quantitative results of the evaluation and qualitative analysis of the 
feedback, allowed us to define specific future work for further improving 
\approach.

We envision \approach being successfully used in three different scenarios: (1) providing automated feedback to the bug reporter at reporting time, prompting a rewrite of the bug report; 
(2) providing useful information (\eg the missing S2Rs) to the developers attempting to reproduce the bug; and (3) supporting automated approaches for test case generation (\eg Yakusu \cite{Fazzini2018}).


\begin{figure}[]
	\includegraphics[width=1.\columnwidth]{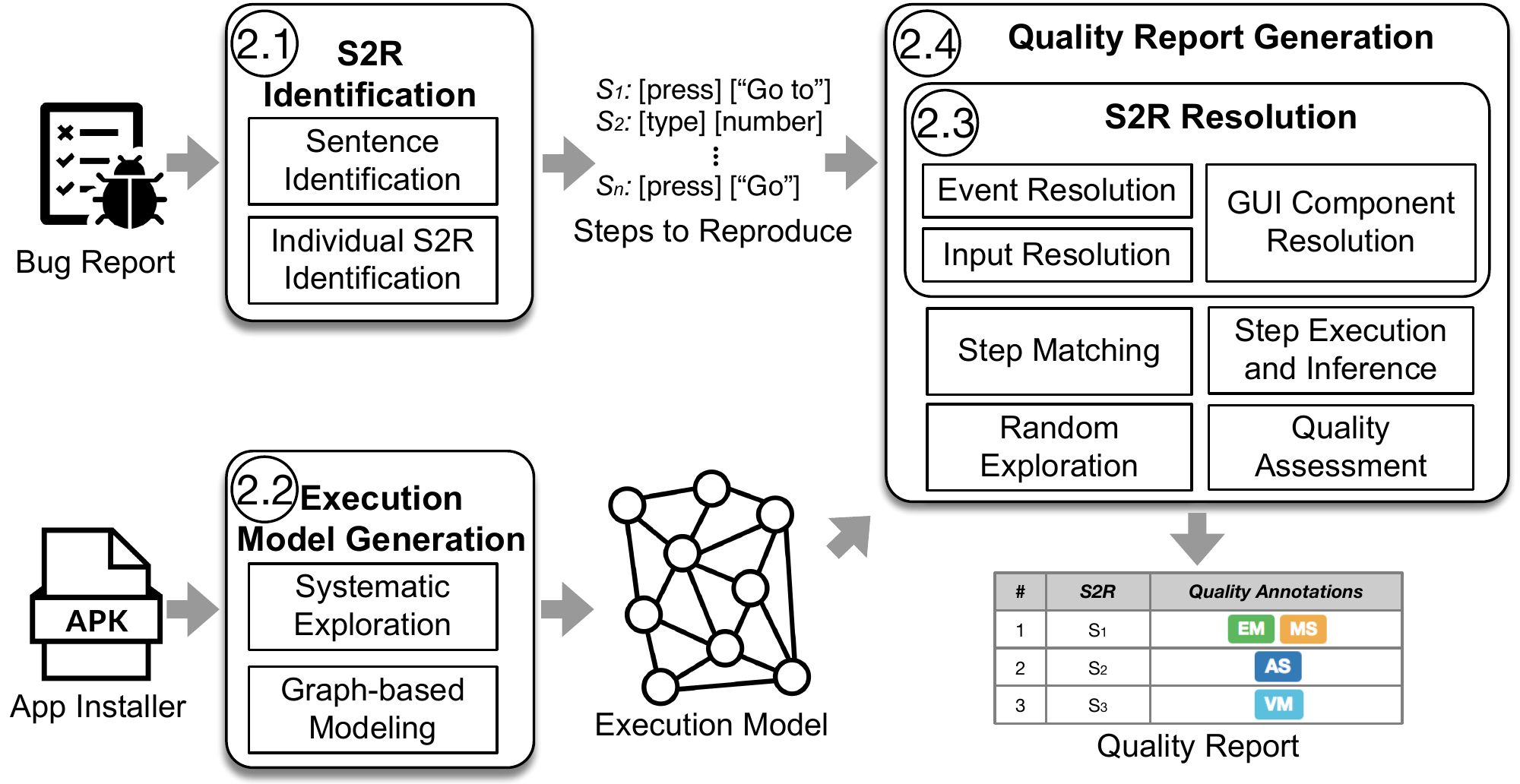}
	\caption{\approach's workflow and main components.}
	\label{fig:euler}
	\vspace{0.3cm}
\end{figure}

\section{Assessing S2R quality}
\label{sec:approach}

We describe \approach (\approachlong), an approach that automatically identifies and assesses the 
quality of the steps to reproduce (S2Rs) in bug reports. In this paper, we 
focus on bug reports for GUI-based Android apps, yet \approach can be adapted 
to work for other platforms.
The input of \approach is the textual description of a bug report and the 
executable file of the Android application affected by the reported bug. The 
output is a Quality Report (QR), which contains a set of Quality Annotations 
(QAs) for each S2R, 
automatically identified from the bug description. The QAs are described in 
Table \ref{tab:s2r_taxonomy} of Sec.~\ref{subsec:quality_annotations}. Figure 
\ref{fig:euler} shows \approach's main components and workflow, which are 
described in the following subsections.

\vspace{-0.2cm}
\subsection{Identifying S2Rs}
\label{subsec:s2r-ident}

The first step in \approach's workflow is the automated identification of 
sentences describing S2Rs. Then, 
\approach performs a grammatical analysis on these sentences to identify individual 
S2Rs. The output is a list of individual S2Rs identified from the bug report.

\subsubsection{Identification of S2R Sentences}
\approachtwo identifies S2R sentences in a bug report using 
a neural sequence labeling model~\cite{Huang2016LSTM,Lample2016neural}, which 
contains the following components:

\textbf{Model Input.}
The model input consists of paragraphs in the bug report. Each 
paragraph is a sequence of sentences, and each sentence  is a sequence of 
words. As there are dependencies between S2R sentences (\ie often they 
appear in sentence groups), we use the Beginning-Inside-Outside (BIO) tagging 
approach \cite{Ramshaw1999NLP}, where for each sentence in a paragraph, we 
assign: (1) the  label [B-S2R] if the sentence begins a S2R description; (2) the 
label [I-S2R] if the sentence is inside the S2R description; or (3) the label [O] 
if it is outside (\ie not part of) the S2R description.

\textbf{Word Representations.} 
We represent each word by concatenating two components: word and 
character embeddings. We use pre-trained word vectors
 from a corpus of 819K bug reports, collected from 358 open source projects, to 
 capture word-level representations. In order to 
handle vocabulary outside of this corpus, we use a one-layer Convolutional 
Neural 
Network (CNN) with max-pooling to capture character-level 
representations~\cite{ma2016end}. 
We model word sequences in a sentence by feeding 
the above word representations into a Bidirectional Long 
Short-Term Memory (Bi-LSTM), which has been shown to outperform alternative structures 
\cite{neuralseq}.
The hidden states 
of the forward/backward LSTMs are concatenated for each word to obtain the 
final word sequence representation. 

\textbf{Sentence Representations.} 
As suggested 
by Conneau \etal~\cite{sentembed}, we adopt the simple (yet effective) approach 
of averaging the vectors of words composing a sentence for capturing 
sentence-level properties. We represent each sentence by 
concatenating the averaged word representations from the previous step and a 
one-hot feature vector 
that encodes the discourse patterns inferred by 
Chaparro \etal \cite{Chaparro2017-2}, which capture the syntax and
semantics of S2R descriptions as well as sentences describing the system's 
observed behavior (OB) and expected behavior (EB). 

\textbf{Inference Layer.}
In order to model label dependencies, we use a Conditional Random Field (CRF) for 
inference instead of classifying each sentence independently. CRFs have been 
found to outperform alternative models \cite{neuralseq}.
The output of 
the inference layer is a label for each sentence where the [B/I-S2R] labels 
indicate a S2R sentence and the [O] label indicates a non-S2R sentence.
Section \ref{subsec:approach-implementation} details the model
implementation, training, and evaluation.

\subsubsection{Identification of Individual S2Rs}

Once the S2R sentences are identified, \approach uses dependency 
parsing~\cite{Manning2014} to determine the grammatical relations between the 
words in each sentence and extract the individual S2R from them. \approach 
utilizes the 
Stanford CoreNLP toolkit \cite{Manning2014} for extracting the grammatical 
dependency tree of S2R 
sentences. This tree varies across different types of sentences (\eg conditional, imperative, 
passive voice, \etc). Therefore, \approach implements a set of algorithms that  
extract the relevant terms from the dependency trees of each 
sentence type. 


An individual S2R complies with the following format:
\begin{quote}
\texttt{\small \act \obj \prep \objtwo}  
\end{quote}
\noindent where the \act is the operation performed by the user 
(\eg tap, minimize, display, \etc), the \obj is an ``entity'' directly affected 
by the \act, and \objtwo is another ``entity'' related to the \obj by the   
\prep. An ``entity'' is a noun phrase that may represent numeric and 
textual system input, domain concepts, GUI components, \etc A S2R example is: 
\emph{``[create] [entry] [for] [purchase]''}.


We illustrate \approach's algorithm to identify individual S2Rs from 
conditional S2R sentences. The bug 
report \#256 \cite{gnucash_bug} 
from  GnuCash~\cite{gnucash} contains the 
conditional S2R sentence: \emphdblquote{When I create an entry for a 
purchase, the autocomplete list shows up}. 
To extract the  S2R from the parsed grammatical tree, \approach first
locates the adverb \emphquote{When} that is the adverbial modifier 
(\emph{advmod}) of the verb (\emphquote{create}). Then \approach verifies the 
existence of an adverbial clause modifier 
(\emph{advcl}) between the verb and its parent word. Next, \approach captures 
the verb (\emphquote{create}) as the S2R's \act. 
Otherwise, 
the sentence is discarded, as it does not follow the grammatical structure of 
a conditional sentence. Next, \approach 
locates the nominal subject (\emph{nsubj}), 
in this case, the word \emphquote{I}. 
\approach 
captures 
the \obj by locating the verb's direct 
object (\emph{dobj}), \emphquote{entry} in the example, and
identifies the 
nominal modifier of the direct object 
(\emph{nmod:for}) as the \objtwo, and the \prep of the 
nominal 
modifier (\emph{case}). 
The resulting S2R is: 
\emph{``[create] [entry] [for] 	[purchase]''}.

The final result of the S2R identification is a sequence of S2Rs extracted from 
the bug report, $S2Rs = \{s_1, s_2, s_3, ..., s_n\}$. The sequence order is 
determined by the order in which the S2Rs appear in the bug description, from 
top to 
down and left to right, except for a few cases, such as ``I do \textit{x} after 
I do \textit{y}", where the order is right to left.

\subsection{Execution Model Generation}
\label{subsec:execution-model}

\approach's quality assessment strategy is based on an execution 	
model
that captures sequential GUI-level application interactions and the 
application's response to those interactions.

In its current implementation, \approach utilizes a modified version of 
\crashscope's \textit{GUI-ripping Engine} \cite{Moran2016,Moran:2017} to 
generate a database of application execution data in the form of sequential 
interactions. This Engine is an automated system for dynamic analysis of 
Android applications that utilizes a set of systematic exploration strategies 
and has been shown to exhibit comparable coverage to other automated mobile 
testing techniques~\cite{Moran2016}. A detailed description of the engine can be found in Moran \etal's previous work~\cite{Moran2016,Moran:2017}.

\approach's next task is the generation of a graph that abstracts the sequential
execution database produced by \crashscope's Engine. The granularity of states 
in this
graph is important, as it will serve as an index for matching the identified
natural language S2Rs with execution information.  For instance, if the graph
were built at the activity-level (meaning that each activity recorded by
\crashscope represents a unique state in the graph), then there is potential for
information loss, as the GUI-hierarchy of a single activity may change as a
result of actions performed on it \cite{Baek2016}.  

To avoid information loss, 
\approach generates a directed graph $G = (V, E)$, 
where $V$ is the set of unique \textbf{application screens} with complete GUI 
hierarchies, and $E$ is a set of \textbf{application 
interactions} performed on the screens' GUI components. In this model, two 
screens with 
the same number, type, size, and hierarchical structure of
GUI components are considered a single vertex. $E$ is a set of  
unique tuples of the form $(v_x, v_y, e, c)$, 
where $e$ is an application event (\eg tap,
type, swipe, \etc) performed on  a GUI component $c$ from screen $v_x$, and 
$v_y$ is 
the resulting screen right after the interaction execution.
Similar execution models have been proposed in prior research 
 on mobile app testing \cite{Zaeem2014}.  
Each edge stores additional information about the interaction, such as the 
data input (only for \textit{type} events) and the 
interaction execution order dictated by the systematic exploration. 
The 
graph's starting node has one outgoing interaction only, which 
corresponds to the application launch. A GUI component is uniquely represented 
by a type (\eg
a button or a text field), an identifier, a label (\emphquote{OK} or
\emphquote{Cancel}), and its size/position in the screen. Additional 
information about a component is stored in the graph, for example, the 
component description given by the developer and the parent/children 
components. 

\subsection{S2R Resolution}
\label{subsec:step-match}

\approach needs to identify the application 
interaction that most-likely corresponds to a S2R
(\aka \textit{step resolution}). Given a S2R 
$s$ and 
program state $v_x$ (\ie graph vertex or screen), \approach determines the 
most 
likely interaction $i=(v_x, v_y=null, e, c)$ for $s$, where $e$ is an event 
performed on component $c$ from the screen $v_x$.  For \textit{type} events 
(\ie text entry events), \approach identifies the input value specified by $s$, if any. 
\textit{Step resolution} can fail to resolve the 
interaction for $s$. In that case, the result is either a mismatch (\ie $s$ 
does 
not match a possible interaction in the current screen) or a multiple-match 
(\ie $s$ matches multiple events or screen components).

\subsubsection{Event Resolution}
\label{sec:step-match:event_resolution}

The first step in the \approach's step resolution workflow is determining the 
event $e$ that a S2R refers to. \approach supports the following Android 
events: tap, long tap, open app, tap back/menu 
button, type, swipe up/down/left/right, and 
rotate to landscape/portrait orientation.

First, \approach finds the \textit{action group} that the \act from the S2R  
corresponds to. An \textit{action group} is a category for verbs having a 
similar meaning, used to 
express an app interaction. \approach finds 
the action 
group by matching the \act's lemma to the lemma of each verb in the group. 
\approach supports six action groups, namely OPEN, LONG\_CLICK, CLICK, SWIPE, 
TYPE, and ROTATE. Each group has a set of verbs (\eg edit, input, enter, 
insert, \etc for TYPE).  We defined the groups by analyzing the 
vocabulary used in the bug reports and applications used by Moran \etal 
\cite{Moran2015,Moran2016}.

When the \act maps to multiple action groups, \approach 
resolves the correct group by analyzing the \obj and \objtwo from the S2R (\eg 
by identifying GUI-component types in them or matching these to 
screen components using the matching algorithm described in Sec.  
\ref{sec:component_resolution}). Only the groups TYPE, CLICK, and ROTATE 
have common verbs. 
If \approach fails to disambiguate 
the action group, then it flags the S2R's \act as matching multiple 
events and saves the corresponding action groups 
for providing user feedback.

If the \act does not match an action group, then the 
verb is likely to refer to a generic interaction or an application 
feature (\eg ``\textit{[create] [purchase]''}).
In this case, \approach assumes the \act is
expressed in the properties of a GUI component (\ie its
ID, description, or label). Then, \approach attempts to resolve a 
GUI component that matches the whole S2R or the \act, by using the matching 
approach define in Sec. \ref{sec:component_resolution}. If there is 
a matched component, the action group is determined as CLICK (if the component 
is tappable), as LONG\_CLICK (if the component is long-tappable), or TYPE (if 
the component is type-able). Otherwise, the event resolution process fails with 
an event mismatch result.

Once the action group is determined, \approach proceeds with translating such 
a group into an event. The OPEN action group is translated as an 
`open app' event if the \obj matches `app', the 
current 
app name, or a synonym (\eg `application'). 
Otherwise, it is resolved 
as a `tap' event. The CLICK group is translated as a `tap back button' 
event,  if the \obj or \objtwo contains the terms `back', `leave', or related 
terms, and as a `tap menu button' event, if the \obj or \objtwo contains the 
terms `menu', `more options', `three dots', \etc Otherwise, it translated as a
`tap' event. The rest of the action groups are translated to their 
corresponding event (\eg TYPE as `type'). We also use keywords to determine the 
direction of swipes and rotations (\eg `landscape', `portrait', `up', `right', 
\etc).

All the keywords mentioned in this section are based on our 
experience with Android apps and the analysis of  Moran \etal's bug reports and apps 
\cite{Moran2015,Moran2016}.

\subsubsection{GUI Component Resolution}
\label{sec:component_resolution}

The next step in \approach's step resolution workflow is determining the GUI 
component in the current screen that the event should perform on, according to 
the S2R. This step is completed only for
tap, long tap, tap 
on menu button, and type events, as they are the only ones that require a 
component. 
Before describing how the GUI component resolution works, we describe the 
base algorithm used to match a textual sequence (\ie a query) to a GUI 
component.

\textbf{Matching algorithm.} The algorithm's input is a 
textual query $q$ (\ie a sequence of terms), a list of GUI components $GC$ 
(sorted in the 
order of appearance in a screen), and the
application event $e$ identified from the S2R. The output is 
the GUI component (from $GC$) most relevant to the query. The 
relevancy is determined by a set of heuristics and a scoring mechanism based on 
textual similarity. The algorithm comprises the following steps:
\begin{enumerate}
	\item If $q$ contains terms referring to the application or device screen (\eg 
	`screen', `phone', \etc), then the first non-tappable component of the current 
	screen (from top to down) is selected and returned as the most relevant 
	component.
	
	\item If $q$ contains terms that refer to a component type, such as \textit{text 
		field} 
	or \textit{button}, then \approach checks if there is 
	only one component in $GC$ of that type (the first type found in 
	the query). If that is 
	the case, then the algorithm selects and returns such a component as the most 
	relevant component.
	
	\item If $q$ does not contain any terms related to component types, then 
	\approach computes a similarity score between $q$ and each component $c$ from 
	$GC$ and selects a set of candidates most-relevant to the query. The similarity
	score is computed as: 	
	\begin{equation}
	\label{eq:similarity}
	similarity(s_{1},s_{2})=\frac{\left|LCS(s_{1},s_{2})\right|}{avg\left(\left|s_{1}\right|,\left|s_{2}\right|\right)}
	\end{equation}
	\noindent where $s_1$ 
	and $s_2$ are 
	two term sequences, $LCS(s_{1},s_{2})$ is the Longest Common Substring between 
	the sequences at term level (as opposed to character level), and 	
	$avg\left(\left|s_{1}\right|,\left|s_{2}\right|\right)$ is the average length 
	of both sequences. If any of the sequences is empty then the 
	similarity score is zero. If two sequences are exactly the same, then 
	the score is 1 (\ie the maximum score), otherwise, the score varies from 0 to 
	1. 
\end{enumerate}

The similarity score accounts for common terms between the sequences 
and the order in which they appear. The order is important because the matching 
process should be as precise as possible for producing an accurate S2R quality 
assessment.  Before computing the similarity, \approach applies lemmatization 
to the input 
word sequences (using the Stanford CoreNLP toolkit \cite{Manning2014}).

The similarity between $q$ and each component $c$ in $GC$ is taken from the 
similarity computed between $q$ and the component label, description, 
and id, in that order. Specifically, the first non-zero similarity score 
obtained from these sources is taken as the similarity between $q$ and $c$. 
Only the components whose similarity with $q$ is 0.5 or greater are considered 
similar to the query, yet \approach recommends candidates in the order of their 
similarity score, with the highest first.

From the candidate list, \approach determines the component that 
is most relevant to the query. There are three cases to consider:
\begin{enumerate}
	\item There is one candidate. \approach returns such 
a component and the matching algorithm ends. 

\item There is more than one candidate. To determine the most-relevant component, 
\approach executes a set of heuristics.
For each component, if its type is Layout and it has only one child in the 
GUI hierarchy, then the child is returned and the process ends.
 If none of the candidates 
satisfy the condition above, but all candidates are of the same type (\eg text 
fields), then the component with the highest similarity score is 
returned. Otherwise, \approach analyzes the candidates 
with respect to the event $e$. If $e$ is a 
typing event, and there is one text field among the candidates, then field is 
returned. 
Otherwise, if $e$ is a `tap' or 
`long tap' and there is only one button among the candidates, then \approach 
returns such a component. Otherwise, the algorithm ends   
with a multiple-match result and the candidates are saved for providing the 
quality feedback to the user. 

\item There are no candidates. \approach 
reformulates the query following a query replacement approach, where a set of 
predefined synonyms for query terms are used as new queries. 
If there are no synonyms for the query terms, then the algorithm stops and 
returns a mismatch result. Each query is executed and if any matches a 
component, then it is returned. Otherwise, the process ends with a mismatch.
\end{enumerate}

\textbf{Query Formulation and Component Resolution.} 
 \euler uses 
the S2R constituents as queries, depending on the identified event $e$ for a 
S2R. These queries are executed using the matching algorithm to find the GUI 
component that the S2R most likely refers to.

For the `tap', `long tap', and `tap on menu button' 
events, the first formulated and 
executed query is the 
entire 
S2R (\ie the concatenation of the S2R's \act, \obj, and \objtwo). If the 
matching algorithm fails to return a component, only \obj or 
\objtwo are executed as queries. In both cases, if the \act 
corresponds to a verb that means ``selecting'' (\eg ``select", 
``choose", ``pick", ``mark", \etc), then only checkable or pickable components 
(\eg drop-down lists or check-boxes) in the current screen 
are used as search space. The \objtwo-based query 
is executed only if the \obj-based one fails. If both queries fail, then 
the query ``\act + \obj'' is reformulated and executed. If any of these queries 
fail to match a GUI component, then the step 
finishes with either a mismatch or a multiple match, depending on the last 
matching result obtained.

For \textit{type} events, \euler considers the following S2R cases:
\begin{enumerate}
	\item A S2R with a literal in \obj, a non-literal in \objtwo, and the 
	\prep is one of the following: ``on'', ``in'', 	``into'', ``for'', ``of'', 
	``as'', \etc For these cases, the \objtwo is used as query. For example, 
	for the S2R \textit{``[enter] [`10'] [on] [price]''}, the term 
	\textit{price} is used as query.
	\item A S2R with a non-literal in \obj, a literal in \objtwo, and the 
	\prep is one of the following: ``to'' or ``with''. Then, the 
	\obj is used as query. For example, for the S2R \textit{``[set] 
	[price] [to] [10]''}, the term \textit{price} is used as query.
	\item A S2R where the \obj is a literal and the \prep and \objtwo 
	are null 	(\eg \textit{``[enter] [`10']''}), \euler selects and 
	returns the focused component in the current screen, if any.
\end{enumerate}
In any case, the resolution process ends with either a resolved GUI 
component or a mismatch/multiple-match result.

\subsubsection{Application Input Resolution} 
\label{sec:input_resulution}

For \textit{type} events, \approach extracts 
the input values from the \obj or \objtwo. Specifically, \approach identifies 
literal values or quoted text. If the input value is missing or generic (\ie 
not a literal or ``text''), then \approach generates a numeric input value 
from a counter (a simple, yet effective approach).

\subsection{Quality Report Generation}
\label{subsec:quality_annotations}

\approach's S2R quality assessment algorithm  receives as input the identified 
S2Rs from the bug report and the system execution graph $G$. The 
output is a Quality Report (QR), providing an assessment and feedback for each 
S2R. 
The algorithm comprises four major steps: (1) step matching; (2) step execution 
and inference; (3) random application exploration; and (4) quality assessment.

\subsubsection{Step Matching}
\label{subsub:step-matching}

\approachtwo attempts to match the S2Rs with application states and 
interactions. 
Starting with the first identified S2R, \approach resolves an
interaction using a set of screens from the graph. First, \approach 
verifies if the first S2R corresponds to an `open app' interaction. If 
it does, then \approach marks the S2R as analyzed and proceeds to
the next S2R. Otherwise, \approach builds the interaction. Either way, 
\approach executes an `open app' event, and the 
target state from this interaction is marked as the current execution state. 
\approach makes sure that the current state corresponds to the screen shown 
on the device.

Starting from the current state, \approach traverses the graph in a 
depth-first manner until $n$ levels have been reached.
\approach performs step resolution on each state (Sec. \ref{subsec:step-match}). The result is a 
set of resolved interactions for the S2R on the selected states. If 
the S2R resolution fails for these states (either with a 
mismatch or a multiple-match result), then it means that either: (1) more 
states in the graph need to be inspected, hence, the parameter $n$ should 
increase; (2) there are app states 
uncovered by the systematic exploration (\ie not present in the execution 
model); or (3) the S2R is of low-quality. The parameter $n$ needs to be 
calibrated per each app. \approach discovers additional app states via random app exploration (Sec. 
\ref{subsub:random-exploration}).

Ideally, only one interaction is resolved for the S2R (\ie on one state 
only). However, it is possible to resolve multiple interactions, each one on 
different app states. This is due to variations in 
the states resulting from different interactions. For example, when 
providing various app inputs, one screen could have a slightly different 
GUI hierarchy. The resolved interactions are matched against the interactions 
from the graph, by matching their source state $v_x$, the event 
$e$, and the component $c$. If a pair of interactions match on these 
properties, then they are considered to be the same interaction. The matching 
returns a set of 
interactions from the graph that match the resolved ones. If this set is empty, 
then it means that the resolved interactions were not covered by the systematic 
exploration approach, and \approach assumes they are new interactions in 
the graph. 
\approach proceeds with selecting the most 
relevant interaction that corresponds to the S2R, by selecting the one whose 
source state is the nearest to the current execution state in the graph. In 
particular, \approach computes 
a relevant $score = 1 / (d + 1)$, for each 
interaction, where $d$ is the distance, in terms of number of levels apart in 
the graph, between the current state and the source state of the interaction. 
\approach selects the interaction with the highest $score$ as the one that 
matches the S2R. This decision is made to minimize the number of steps required 
for reaching the state where the interaction is executed, as described below.

\subsubsection{Step Execution and Inference}
\label{subsub:step-exec-n-inference}

Each identified interaction from the graph is executed in the device.  
Any new 
application screens/interactions are added to the graph during the execution. 

The identified interaction in the graph for a S2R could be 
located in a state far away from the current state. This means that \approach 
needs to execute intermediate interactions for reaching the state where the  
interaction is executed on. There may be more 
than one way to reach such a state. Therefore, \approach 
selects the shortest path between the current state and the state where the 
interaction occurs. The interactions in the shortest path are assumed by 
\approach 
as inferred steps, missing in the bug report. \approach 
executes each one of the interactions in the shortest path.
At each state, 
\approach determines the enabled components in the device screen and only the 
interactions to such components are executed, in the order that they were 
executed by the systematic exploration approach or the 
current execution. All the interactions executed correspond to the list of 
inferred interactions or missing steps in the bug report.

\subsubsection{Random System Exploration}
\label{subsub:random-exploration}

As mentioned before, if the step resolution fails for all the inspected states, 
then it means that the systematic app execution approach (Sec. \ref{subsec:execution-model}) failed to discover app 
states/screens. To address this issue, \approach performs a random app 
exploration, starting from the current app screen 
(shown on the device). The goal is to discover additional app 
states that could lead to successfully resolving the interaction for a S2R. 
To do so, \approach identifies the components (different than Layouts and List 
Views) 
that have not been executed in 
the current screen, and randomly selects and executes one clickable component 
from this set.

The random exploration is performed iteratively $y$ times. At each iteration, 
$x$ interactions are executed, unless there are no components left to interact 
with  
in the current screen. Right after each iteration, \euler 
updates the graph, the app is restored to the state before the random 
execution, and the S2R matching, execution, and inference are 
performed again on the graph's new version. If the S2R is matched against 
the graph (Sec. \ref{subsub:step-matching}), 
then no more iterations are executed. Else, the random 
exploration process continues.

\subsubsection{Quality Assessment}
\label{subsub:quality-assessment}


\begin{table}[]
	\caption{Quality annotations for the S2Rs in bug reports }
    \label{tab:s2r_taxonomy}
	\centering{}%
	\vspace{-0.8em}
	\small
	\begin{tabular}{l}
\hline
\textbf{High Quality (HQ):}   \\      
A step that precisely matches an application 
interaction \\ \hline
\textbf{Low Quality (LQ) -  Ambiguous Step (AS): }      \\
 A step that matches more than one GUI component or event   \\ \hline
\textbf{Low Quality (LQ) - Vocabulary Mismatch (VM):}  \\
A step that does not match any application interaction  \\ \hline
\textbf{Missing Step (MS):}  \\
 A step required to reproduce the bug, not described in the bug report \\ 
 \hline
\end{tabular}
\vspace{0.2cm}
\end{table}

\approachtwo assigns a set of Quality Annotations (QAs) to each S2R. The QAs are defined in Table \ref{tab:s2r_taxonomy}.  
If the S2R is resolved/matched against the execution model successfully, 
then \approach labels the S2R as High Quality (HQ) - \aka Exact Match~(EM).

If there are inferred application interactions between the previous S2R and the 
current one, then the current S2R is labeled with Missing Steps (MS). The 
inferred steps are attached to the annotation for informing the reporter about 
them. The feedback given to the users is that there are application 
interactions missing in the bug report that should be executed before 
the current S2R. Note that the MS annotation does not indicate a problem about 
this S2R but about the entire list of S2Rs.

If a S2R is not resolved in any of the graph states because of a 
multiple-component or -event match, then it is labeled as an Ambiguous Step 
(AS). 
The feedback given to the users is that either the S2R's \act 
corresponds to multiple events, or the \obj or \objtwo match 
multiples GUI components. Examples of matched events or components are shown 
to the user.

If a S2R is not resolved in any of the graph states because of a mismatch of 
the S2R with the application, then it is labeled with Vocabulary Mismatch (VM). In the feedback given to the user, \approach specifies the 
problematic vocabulary from the S2R constituents (\ie the \act, \obj, \objtwo, 
or any combination of these).

\begin{figure}[h]
	\begin{centering}
		\includegraphics[width=1\columnwidth]{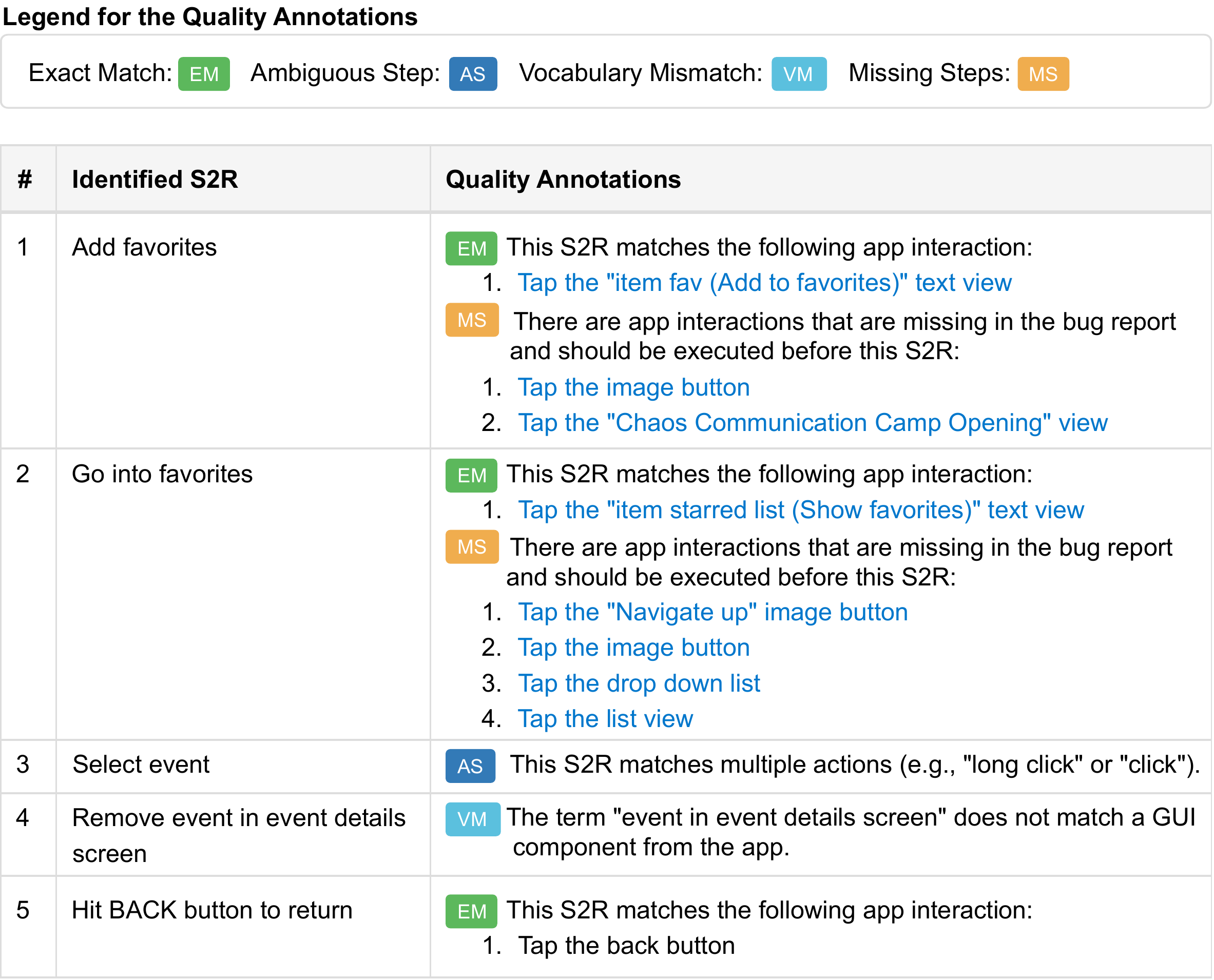}
		\par\end{centering}
	\caption{\approach's Quality Report for Schedule \#154 \cite{schedule_154}.}
	\label{fig:QR}
\end{figure}

\approach generates a web-based Quality Report 
(QR) with the quality assessment for the S2Rs in a bug report, containing the 
feedback described above (Fig.~\ref{fig:QR}). The user can click on 
the matched/inferred interactions to open a pop-up window  showing a screen 
capture of the app, highlighting the GUI component being interacted with.

\section{Empirical Evaluation}
\label{sec:study}

We conducted an empirical evaluation to determine how accurately
\approach identifies and assesses the quality of S2Rs in bug 
reports, and to understand the perceived usefulness, readability, and 
understandability of the information included in 
\approach's Quality Reports (QRs). We aim to answer the following research 
questions (RQs):

\newcommand{\rqone}{What is the accuracy of \approach in identifying and assessing the quality 
of the S2Rs in bug reports?} 
\newcommand{\rqoneone}{To what extent is \approach able to correctly generate S2R quality annotations?}
\newcommand{\rqonetwo}{To what extent is \approach able to identify missing S2R?}
\newcommand{\rqtwo}{What is the perceived usefulness and quality of the information provided in \approach's quality reports?}

\begin{quote}
{\bf RQ$_1$} \emph{\rqone}	
\end{quote}

\begin{quote}
	{\bf RQ$_2$} \emph{\rqtwo}
\end{quote}

The answer to RQ$_1$ will inform us on improvements to \approach's accuracy. 
RQ$_2$ will inform us on the presentation and perceived usefulness of the 
information provided by the QRs.

In order to answer the RQs, we selected a set of bug reports  
(Sec.~\ref{subsec:bug-sampe}), collected human-produced reproduction 
scenarios for them  (Sec. \ref{subsec:bug_scenarios}),  
used \approach to identify and assess the quality of each S2R (Sec. 
\ref{subsec:approach-implementation}), and asked external 
evaluators to assess \approach's QRs (Sec. \ref{subsec:methodology}). We 
analyze the resulting evaluation data and 
answer the RQs using a set of evaluation metrics, defined in Sec. 
\ref{subsec:metrics}.

\subsection{Bug Report Sample}
\label{subsec:bug-sampe}

We used 24 bug reports from six Android apps 
\cite{Moran2015,Moran2016}: (1) \textit{Aard Dictionary}, a dictionary and 
Wikipedia reader \cite{aardict}, (2) \textit{Droid Weight}, a body weight 
tracker \cite{droidweight}, (3) 
\textit{GnuCash},
a finance expense manager \cite{gnucash}, \textit{Mileage},  a vehicle mileage 
tracker \cite{mileage}, 
\textit{Schedule}, a conference scheduler \cite{schedule}, and (6) \textit{A 
Time Tracker} \cite{atimetracker}. The apps were selected to 
cover different domains,  as well as involve 
multiple events (\eg taps, types, swipes, \etc) for using their functionality. These apps are also well-studied, having been utilized in several past works on mobile testing and bug reporting~\cite{Moran2016,Moran2015}.

We collected the entire set of issues (\ie 785, excluding pull requests) from 
the issue trackers of the six apps. We randomly sampled 56 issues (\ie about 
10\% of the data for each app except GnuCash, which had 
the largest issue set, and its sample amounts to 5\% of the issues). We read 
the issues and discarded 32, which 
correspond to new feature requests, enhancements, \etc, or bug reports with no 
S2Rs included. 
The remaining 24 issues 
correspond to bug reports, and out of these, 20 describe reproducible bugs and 
4 describe non-reproducible bugs. 
The reports describe different 
types of bugs, namely crashes (5 reports), functional problems (14 
reports), and look-and-feel problems (5 reports). The reports include 88 S2Rs 
total, 3-4 S2Rs per report on avg., with min. 1 and max. 8. 

We manually inspected the 88 S2Rs and estimated that 68 steps are of 
high-quality, 16 are 
ambiguous, and four use unexpected vocabulary, while there are many missing 
steps.

\subsection{Ideal Reproduction Scenarios}
\label{subsec:bug_scenarios}

In order to assess the quality of the S2Rs from the sampled reports, we need a 
baseline: the ideal list of S2Rs (\aka ideal
reproduction scenarios). To build the scenarios, we asked six
graduate students to reproduce the reported bugs by 
following the S2Rs provided in the reports. Each bug was reproduced by 
two students. For each bug report, a student had to (1) (re)install the 
buggy version of the app on an Android emulator, and (2) try to 
replicate the reported bug, while writing (in a spreadsheet) each 
specific step followed. In some cases, the students attempted to replicate the 
bug more than once. On each attempt, they annotated the 
detailed reproduction sequence, including any missing steps in 
the bug report. 
In most cases, the students succeeded reproducing the reported bug on their second attempt (for 
the reproducible bugs). The scenarios across the two students per report 
were highly similar, if not the same. We found only small variations in the 
scenarios for a single bug (\eg 
input values, or cases such as \textit{tap back button} vs. \textit{tap  
cancel button}). 
  
From the collected reproduction scenarios, we created the ideal reproduction 
scenario (\ie the ideal S2Rs) for each bug report,  which includes the set of 
missing steps in the report and the correspondence for each app 
interaction/step (in the scenario) with the S2Rs from the report. For each reproducible bug, 
we selected the steps that are more clearly-written, among the submitted scenarios. When necessary, we decomposed the 
steps into atomic app interactions and added step details (\eg the 
location of the GUI-components). We also normalized the 
vocabulary (\eg `hit' or `press' are changed to `tap'). For each 
report describing a non-reproducible bug, we selected the two most similar 
scenarios to the bug report scenario, and performed the same 
normalization procedure.

\subsection{\approach Implementation and Calibration}
\label{subsec:approach-implementation}

We implemented \approach's S2R identification component by adapting 
the NCRF++ toolkit \cite{ncrf}.  We trained the word embeddings with dimension 
200 on 819K bug reports collected from 358 open source projects using 
the \textit{fastText}'s skip-gram model implementation \cite{fasttext}. We used  data from 
Chaparro \etal \cite{Chaparro2017-2} to train the model, using 
data from GUI-based systems only.  The character embedding layer 
consists of one convolution layer with kernel size of 3. The size of the 
character vectors is 50, the size of bi-LSTM vectors is 40, and the size of the 
discourse patterns vector is 154 \cite{Chaparro2017-2}.

For learning, we use a mini-batch of size 4 using stochastic gradient descent 
with a 0.05 decayed learning rate to update the parameters. The learning rate 
is 0.015. We apply 0.5 dropout to the word embeddings and character 
CNN outputs. We find 
the best hyperparameters by performing a 10-fold cross validation with 
80\%, 10\%, and 10\% of the data for model training, validation, and testing, 
respectively. 
The model is trained for up to 500 epochs, with early stopping if the 
performance (based on F1 score) on the validation set does not change 
within 25 epochs. 
The model achieves 73\% precision and 81\% recall at identifying S2R sentences.

We implemented the remaining \approach components using the Stanford CoreNLP 
library \cite{Manning2014}, Chaparro \etal's implementation of the discourse 
patterns \cite{Chaparro2017-2}, and 
\crashscope \cite{Moran2016}. We used the bug reports by Moran \etal 
\cite{Moran2015,Moran2016} to test our implementation and calibrate the 
parameters.
In particular, \approach executes 3 random exploration 
iterations, with 10 steps each. The depth of graph exploration for the step 
matching is 6 levels from the current program state. These represent 
the best parameters, according to our tests.

\subsection{Methodology}
\label{subsec:methodology}

To address our RQs, we asked human evaluators to assess the 
quality reports generated by \approach.
The study \emph{subjects} (\aka participants) are  six PhD students, one 
business 
analyst, three professors, one postdoc, and one MSc student. The participants 
have 
been selected through direct contacts of the authors, taking into account 
that (i) participants require to have some development experience; and (ii) 
they need to be available for a task of about two hours.

Based on the ideal reproduction scenario, we created a reproduction 
screencast showing how the bug can be reproduced, or, for the 
non-reproducible bugs, how  the sequence of steps could be followed. For each 
bug report, each participant had the following 
 information 
available: (1) the original bug report; (2) the quality report generated by 
\approach; (3) the ideal reproduction scenario; and 
(4) a screencast showing how the bug can be reproduced on a device. Before 
starting the task, we 
instructed the participants in a training session (also made available to them 
through a video), in which we explained the quality annotations and the task to 
be performed. 

We randomly assigned six bug reports to each participant, for which he/she had to 
evaluate the QR; each QR 
is evaluated by three participants. 
The survey questionnaire, implemented through Qualtrics 
\cite{qualtrics}, consists of a demographics section and a 
section for each QRs to evaluate. In the demographics 
section we ask questions about years of experience on (i)  non-mobile app 
development, (ii) mobile app development, (iii) Android app development in 
particular, and (iv) use of Android phone. We also ask approximately how many bug reports
the participant has ever reported.

For each QR, the questionnaire contains two sections. The first section contains, for each S2R, three questions, for answering RQ$_1$:
	\begin{enumerate}
		\item A yes/no question for checking whether \approach correctly 
		identified the S2R (in case of a negative answer, questions (2) and (3) 
		are skipped).
		\item For each annotation produced by \approach for a given S2R, an 
		agree/disagree question aimed at checking its correctness.
		In case the answer was negative, the respondents were instructed to 
		explain their answer in a free-text form.	
		\item In case of missing steps, a third
		(four check-box) question is formulated for assessing 
		whether \approach's suggested list of
		missing steps is: (i) correct; (ii) contains extra steps; (iii) is 
		lacking one or more steps; or (iv) some steps are incorrectly ordered. 
		We ask the respondents to use a free-text 
		form to provide an explanation for their answer.
	\end{enumerate}

The second section of the survey addresses RQ$_2$, by asking:
	\begin{enumerate}
	\item Whether \euler's quality report is easy to read and 
	understand (using a 5-level Likert scale \cite{Oppenheim:1992}).
	\item Whether the quality report is likely to help users to better write 
	bug reports (using a 5-level Likert scale).
	\item Four free-text questions to indicate what information was perceived 
	useful, useless, and what information should be added to or 
	dropped from the QR.
    \end{enumerate}

\subsection{Metrics}
\label{subsec:metrics}

For addressing RQ$_1$, we measure \approach's precision and recall at 
identifying the S2Rs from the bug report by
comparing \approach's output with 
the ideal reproduction scenario (Sec. \ref{subsec:bug_scenarios}).
We also measure the proportion of correctly identified S2Rs  
judged by the participants. Since we involve three participants for 
each bug report, we consider the correctness assessment provided by the 
majority. 

Regarding the QAs for each step, we 
compute, for each QA type (Tab. \ref{tab:s2r_taxonomy}), the proportion of 
annotations 
judged as correct. We consider the assessment of the majority of participants 
requiring at least two positive answers. Note that, in this case, a 
respondent might not have answered question \#2 if she judged the S2R as 
incorrectly identified. 

For MS annotations, we measure the proportion of MS annotations 
suggesting  correct, extra, lacking, and unordered missing steps. 
We also use majority assessment.

To address RQ$_2$, for each bug report we have two questions, expressed 
in a 5-level Likert scale. We compute the cumulative number of 
responses for each of the five levels and we represent them using an 
asymmetric stacked bar chart.

Regarding the free-text questions related to the usability/quality of  the 
QRs information, we categorized the responses using a card-sorting approach 
\cite{cardSorting} and analyzed each category.



\subsection{Results and Analysis}
\label{subsec:results}

\begin{table*}[t]
	\setlength\tabcolsep{2.7pt}
	\centering{}\caption{Accuracy results for \approach's Quality Annotations 
	(QAs).}
	
	\label{tab:rq1-feed}
	\begin{tabular}{l|c|c|c|c|c|ccc|c}
		\hline 
		\multirow{2}{*}{\textbf{App}} & \textbf{\# of}  & \multirow{2}{*}{\textbf{\# 
		S2Rs}} & \multicolumn{1}{c|}{\textbf{\# QAs}} & 
		\multicolumn{1}{c|}{\textbf{\# AS}} & \multicolumn{1}{c|}{\textbf{\# 
		HQ}} & \multicolumn{3}{c|}{\textbf{\# MS}} & 
		\multicolumn{1}{c}{\textbf{\# VM}}\tabularnewline
		\cline{4-10} 
	    & \textbf{Bug rep.}	&  & \textbf{Correct/Tot.} & \textbf{Correct/Tot.}  & 
		\textbf{Correct/Tot.} & \textbf{Correct/Tot.} & \textbf{Not Reported} 
		& \textbf{Extra} & \textbf{Correct/Tot.}\tabularnewline
		\hline 
		Aard Dictionary & 2 & 6  & 5/8 & 0/1 & 4/4  & 1/3 & 0  & 3  & -\tabularnewline	
		A Time Tracker & 5 & 22 &  23/29 & - & 15/18 & 5/8 & 2 & 8 & 3/3 \tabularnewline		
        Droid Weight  & 2 & 6 & 7/7 & -  & 5/5  & 1/1 & 1 & 1 & 1/1  \tabularnewline
	    GnuCash &  9  & 35 &34/53 &1/2&16/26&12/19& 5 & 13 & 5/6 \tabularnewline
        Mileage & 4   & 12 &12/17& 2/2 & 4/6 & 4/5 & 0  & 5  & 2/4\tabularnewline
        Schedule & 2   & 8  &10/10& 1/1 & 5/6 & 3/3 & 0  & 3  & 1/1\tabularnewline
		\hline 
		\multicolumn{1}{l|}{\textbf{Total}} & \textbf{24} & \textbf{89}  & \textbf{91/124} & 
		\textbf{4/6} & \textbf{49/64} & \textbf{26/39} & \textbf{8} & 
		\textbf{33} & \textbf{12/15 }\tabularnewline
		\hline 
		\textbf{\%} & \textbf{-} & \textbf{-} &\textbf{73\%} & \textbf{67\%} & 
		\textbf{77\%} & \textbf{67\%} & \textbf{21\%} & \textbf{85\%} & 
		\textbf{80\%}\tabularnewline
		\hline 
	\end{tabular}
\end{table*}

\tabref{tab:rq1-feed} summarizes the evaluation results of 
\approach's quality assessment and feedback\footnote{Our replication package 
\cite{repl_pack} contains evaluation data and additional results that enable
	the replication of the evaluation. The package includes bug reports, ideal 
	reproduction scenarios,
	identified S2Rs in the reports, \approach's quality reports, study survey, 
	\approach's calibration
	data, and detailed results.}.
It reports 
the 
number of 
S2Rs (identified by \approach) for each 
bug report (3rd column - \# S2Rs), the (correct/total) number 
of quality annotations for all S2Rs (4th column - \# QAs), the (correct/total)  
number of annotations across the quality categories from Table 
\ref{tab:s2r_taxonomy} (5th-10th columns), and the number of MS annotations for 
which 
there are unreported and extra steps in the list of missing/inferred steps (8th 
and 9th 
columns, respectively).

\subsubsection{S2R Identification Results} 
\approachtwo identified 89 S2Rs in the 24 bug reports (Table 
\ref{tab:rq1-feed}).
Only four S2Rs were judged as incorrect, resulting in 96\% overall precision. 
More specifically, the precision is 100\% for 20 bug reports, 
with the exception of four: Aard Dict. \#81 (80\%)~\cite{aardic_81}, 
A Time Tracker \#1 (75\%)~\cite{atimetracker_1}, GnuCash \#471 
(80\%)~\cite{gnucash_471}, and Schedule \#169 (67\%)~\cite{schedule_169}. In 
73/89 (\ie 88\%) answers 
there is  a perfect consensus among the evaluators across bug reports. We 
also found that 
two S2Rs were not identified by \approach (\ie 98\%~recall).

We manually analyzed the four misidentified S2Rs and found that 
the sentences where they were identified from follow the grammatical structure 
of an S2R (\ie conditional, imperative, \etc), but either: (1) they do not 
describe an S2R (\eg ``\textit{Change so the week... is restored}'', from A 
Time Tracker \#1, is addressed to the developer for fixing the bug); (2) they 
indicate an app behavior (\eg ``\textit{when dictionary is being verified}'' 
from Aard Dict. \#81); (3) they are generic actions (\eg ``\textit{When I 
perform these sequences of events}''); or (4) they indicate steps to further 
show how the app correctly behaves  in certain circumstances (\eg ``\textit{It 
shows up again, when you leave the account...}'' from GnuCash \#471). The two 
S2Rs not identified by \approach are misspelled or written using noun 
phrases.

\subsubsection{Quality Assessment Results} 
Table \ref{tab:rq1-feed} shows that (overall) 73\% of the provided QAs/feedback 
were 
considered correct by the evaluators, with a percentage ranging between 67\% 
and 80\% of correct MS and VM annotations, respectively. The participants 
reached a perfect consensus in 56\% of the cases. For 12 bug 
reports, \approach achieves 100\% accuracy. For the remaining 12 reports, 
\approach's accuracy ranged from 0\% to 80\%.
We determined the causes of such performance by manually analyzing 
the participants' answers and \approach's algorithm for those 12 cases, across 
the QA types. 

For two bug reports (\ie Aard Dict. \#104~\cite{aardic_104} and GnuCash 
\#620~\cite{gnucash_620}), \approach 
incorrectly produced two AS annotations (\ie for two S2Rs). According to the 
participants' explanations of their judgment, we found that the 
annotations were confusing to them, specifically, it was not clear which 
components \approach's feedback was referring to. For 
instance, for the Aard Dict. \#104's only S2R: ``\textit{Tap link to another 
Wikipedia 
article}'', \approach produced the AS annotation: ``\textit{This S2R matches multiple 
GUI components (\eg the ``1st Link'' and "2st Link " views)}''. In 
this case, \approach reached a Wikipedia webpage with multiple links having the 
labels shown in the annotation. This webpage was unknown to the participants 
(as it was not shown in the video), hence they did not understand the suggested 
matched components. In addition, we found that the AS annotation produced for 
GnuCash \#620's 
1st S2R: ``\textit{Set the color of an account}'' did not suggest the correct GUI 
component (\ie the color picker in the "creating/editing accounts" 
screens). The cause for such a mismatch lies in the priority that 
\approach gives to resolved interactions from program states closer to the 
current one. One possible 
improvement is to weight in the similarity 
score obtained by the resolved components across multiple program states, 
in such a way that candidates with higher similarity in screens further 
away from the current one are more likely to be suggested.

In six bug reports, \approach incorrectly assessed the quality of 15 S2Rs as  
High quality (HQ), which means that the interactions matched/suggested by 
\approach do not correspond to the S2Rs. We manually analyzed these cases, 
and found four main reasons: (1) the similarity threshold 
defined in Sec. \ref{sec:component_resolution} (\ie 0.5) is too restrictive 
for some reports; (2) the 
similarity used to resolve an S2R to a screen (\ie Formula 
\ref{eq:similarity}) does not account for small term differences between the 
S2R (\ie the query) and GUI components; (3) the synonyms for 
some terms, used to reformulate the query, may 
incorrectly boost the similarity score of unexpected GUI components; and (4) 
the quality of screen information for some applications is low. 

We 
illustrate the first three problems with the report A Time Tracker \#35 
\cite{atimetracker_35}.  The 
first S2R for 
this report was identified as ``\textit{Restore backup}'' and the expected component for 
the S2R is the menu option ``\textit{Restore from backup}'', whose similarity 
to the 
S2R is 0.4 (the LCS is `restore` and the average size of both 
strings is 2.5 - see Formula \ref{eq:similarity}). Because the similarity is 
lower than the threshold, the  component is not returned as a candidate. Next, 
using the predefined query synonyms, \approach 
reformulates the query by expanding the S2R to 
``\textit{Restore back up}'' which returns the menu option ``\textit{Back up 
to 
SD card}'', 
whose similarity to the query is 0.54. In this case, the synonym for backup, 
``back up'', boosted the similarity of the menu option, which was 
returned as most similar to the S2R. 
 To address these problems, we plan to 
improve \approach's similarity formula for cases with 
little term variations, by utilizing shared term 
frequency and how many terms are in-between the shared terms.
To illustrate the fourth problem, 
consider the case of the 5th S2R from GnuCash \#701~\cite{gnucash_701}: 
``\textit{Click `save'}". The 
incorrect matched component for this S2R was the button ``Delete'' from the 
"delete 
account" screen. The component ID 
given by GnuCash developers was ``btn save'', which matches the query. In this 
case, the mismatch could be used as feedback for developers about problems with 
the app screens information. Studying the impact of low-quality app information 
on \approach is subject of future work.

The three S2Rs (from three bug reports) for which \approach incorrectly 
detected a vocabulary mismatch (VM), involve more than one interaction. For 
instance, for the 2nd S2R from GnuCash \#616~\cite{gnucash_616}: 
``\textit{Select export to `Google 
Drive'}'', \approach failed to match ``Google Drive'' because of uncovered 
application states/screen and imprecise S2R parsing and matching.

\subsubsection{Analysis of MS annotations}

We analyze the results for the MS annotations, 
which include the steps inferred by \approach.

The participants reached a perfect consensus in 53\% of the cases with MS 
annotations. 
For six bug reports, \approach incorrectly flagged 13 
S2Rs as having missing 
steps (\ie they were assigned an MS annotation). For the remaining 26 S2Rs (\ie 
66.7\%), 
from 16 bug reports, the MS annotation was correct (\ie indeed there are  
missing steps). For the 13 S2Rs with incorrect MS annotations, all the MSs
 suggested by \approach are unnecessary for bug reproduction. 
For the 26 S2Rs with correct MS annotation, \approach suggested extra MSs 
for 20 of them (\ie 77\%), according to the 
external evaluators. This means that 33 S2Rs, in total, were judged to have 
extra missing steps (see the 9th column of Table \ref{tab:rq1-feed}), which 
represents 85\% of the cases. In addition, for 8 S2Rs total, the list of 
suggested missing steps lacks additional steps (\ie not detected by \approach), 
which represents 21\%. In all MS annotations, the order of the suggested MSs is 
correct, meaning that \approach suggests feasible execution paths. However, in 
all cases, the suggested MSs lack some or have extra steps.

In order to further understand the ability of \approach at inferring and 
detecting missing steps (MSs), we compared the steps suggested by \approach 
against the MSs from the ideal bug reproduction scenarios, and 
computed precision and recall. \approach is designed to favor high recall, because
 it would be easier for a reporter to just select from 
the list of missing steps, the ones she actually did and failed to report, 
as opposed to trying to infer what steps may be missing. Across the 24 bug 
reports, 
\approach inferred and suggested 293 MSs (14 steps per bug report on 
avg.), and 
there are 158 MSs (6.6 steps on avg.) in the ideal reproduction 
scenarios. Our analysis reveals that 92 (4.8 on avg.) suggested MSs are 
correct (\ie true positives), which represents 31\% precision \& 
58\% recall. The results mean that \approach was able to infer more than half 
of the expected MSs.

We analyzed the 13 cases (from 6 bug reports) for which \approach incorrectly 
indicated missing steps, and from the correct MS cases, the 20 cases with extra 
steps. Our 
analysis reveals two main reasons 
for such cases, 
namely, excessive application exploration, and  imprecise S2R 
resolution/matching. Regarding the first limitation, we found that the 
systematic 
and random exploration strategies execute more interactions than needed. 
While this is done by design, trying to uncover as many program 
states/screens as possible, it leads to excessive inferred steps.
 Regarding the second problem, any mismatch 
in the first S2Rs from a bug report can divert \approach's execution, thus 
producing even more mismatches or no matching at all for the remaining 
S2Rs. In the latter 
case, the random exploration takes place, thus producing unnecessary inferred 
steps. We found that the reason for such mismatches comes from the inability of 
the similarity 
scoring formula (\ie Formula 
\ref{eq:similarity}) to match the query with single-term text sequences (from 
the components), and also, from the fact that, in some cases, the random 
exploration is executed late (after the first S2R 
matching fails) and unexpected components, with similar vocabulary to the S2R, 
are returned. Improving the systematic application exploration to uncover as 
many program states as possible may help to alleviate this problem.

Finally, we manually analyzed 10 bug reports for which \approach obtained 
the lowest recall (within the [33\% - 78\%] range) when inferring the expected 
MSs. The main reasons for these cases include: (1) incorrect 
detection of the 
S2Rs' order from the bug report, and (2) failing to handle special S2Rs.
We illustrate the 
first issue with A Time Tracker \#10~\cite{atimetracker_10}. The 
S2R 
sentence ``\textit{If I press the Back button while viewing the Preferences}'' 
implies the S2Rs ``view preferences'' and ``press back button'' are executed in 
that order. \approach failed to identify the correct order in this case, 
provoking to not execute one of the MSs: ``\textit{click OK button}''. One 
exemplar of the second problem is repetitive S2Rs (\eg ``\textit{enter few fill 
ups}'' from Mileage \#53~\cite{mileage_53}), which in its current version, 
\approach does not support.

\begin{figure}[t]
	\centering
	\includegraphics[width=1\columnwidth]{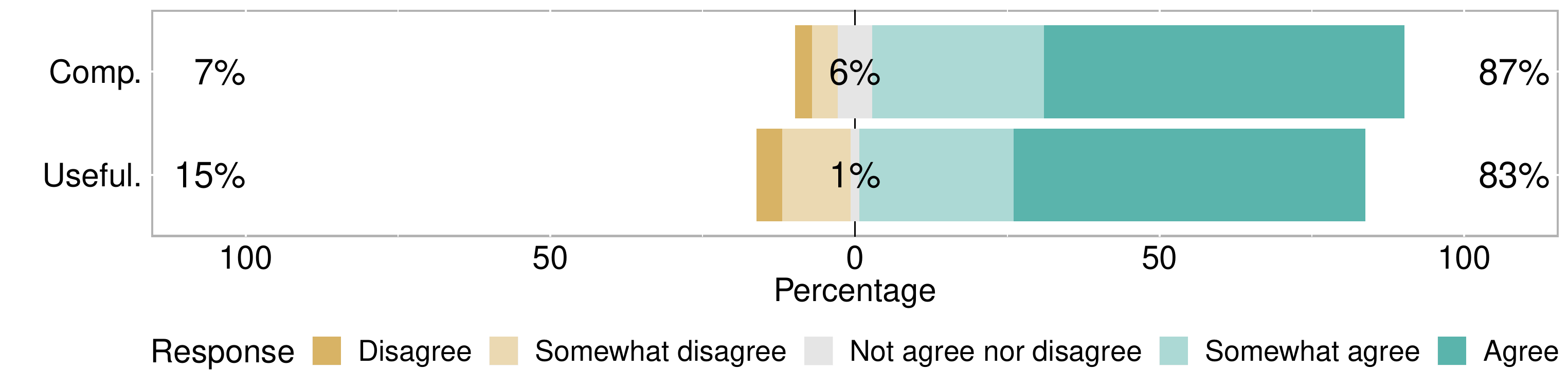}
	\caption {Perceived Comprehensibility \& Usefulness of QRs}
	\label{fig:rq2}
\end{figure}

\subsubsection{Perceived Usefulness} 

\figref{fig:rq2} shows an asymmetric stacked bar chart depicting the perceived 
comprehensibility and usefulness of \approach's quality reports. The 
figure 
shows positive results. In particular, the study participants agree and 
somewhat agree that:
\begin{itemize}
	\item The quality reports are easy to understand (in 59\% and 28\% of the 
	cases, respectively --- 87\% on aggregate).
	\item The quality report can help users write better bug reports (in 58\% 
	and 25\% of the cases, respectively --- 83\% on aggregate).
\end{itemize}

To better understand these results, we analyzed the participants' answers to 
the 
open-ended questions about useful/useless information in the QAs, and 
information that should be added/removed. Our card-sorting analysis 
resulted in the following categories of useful information produced by 
\approach:
\begin{itemize}
	\item Explicit, clear, and fine-grained	S2R feedback. One participant 
	mentions that the matched/suggested ``\textit{S2Rs are pretty descriptive 
	and would guide the user to 	 complete better the bug description}''. 
	Another participant states that the suggested 
	S2Rs ``\textit{are a good example of how to  write steps-to-reproduce}''. 
	Another person acknowledges that ``\textit{Developers/maintainers would 
	find this tool *very* useful for their debugging process}''. Finally, 
	one participant mentions that \approach ``\textit{provides detailed 
	feedback for every single step, not just the overall feedback on all 
	steps}''
	 \item Feedback about incorrect S2R vocabulary. For example,  one 
	 participant	 indicated that the tool correctly ``\textit{flags words 
	 such as `find' and  `fix' that do not directly translate to an app 
	 action}''. Another person notes that \approach detects cases that
	 ``\textit{the user could improve including some 
	 more details in his error report}''. 
    \item Feedback about missing steps. For instance, one participant mentions 
    that the suggested missing steps ``\textit{can 
	 guide the user to list them better}'' in the  report. Another person 
	 mentions that ``\textit{They help avoid the  guessing part 
	 when reproducing the bug}''.
	 \item Screenshots for the matched interactions. Some evaluators considered the 
	 screenshots helpful for ``\textit{identifying the right scenario for 
	 reproducing  the bug}'' and they were found to ``\textit{complement the 
	 description of  each suggested missing step}''.
\end{itemize}

Regarding information that should be added to \approach's quality reports, some
participants suggest that \approach should assess the quality of the 
application observed and expected behaviors, ``\textit{because the user  
described them but they are not clear}''. 
Other participants suggest clearer wording of the S2Rs (\eg ``instead of 
"Tap the `menu save (Export)' text view" >> `Tap Export in menu' ''), and 
visual improvements to the quality report (\eg adding ``\textit{image or some 
sort of representation of `navigation drawer' to help locate the button}'').

The participants also provided feedback about useless/unclear information in 
\approach's QAs that should be improved or discarded. Besides 
incorrect feedback, resulting from \approach's inaccuracy, the participants 
remarked 
that:
\begin{itemize}
	\item Some feedback is unclear. For example, one participant 
	mentions that ``\textit{Some missing steps have strange names}''.	
	 We confirmed that some AS annotations are confusing and found that 
	 the suggested missing steps may be give ``\textit{little information so 
	 the user always needs to click on them and see the image to 
		fully understand the nature of the step}''. \approach phrases the 
	suggested/matched S2Rs based on the app internal data.
	In some cases, this information 
	is not available or may be only clear for the developer (\ie using 
	``btn save'' instead ``Save button'').
	\item Setup application steps are not needed. One participant commented 
	that 
	``\textit{some steps that describe the app initial configuration... are not 
		needed to reproduce the bug}''. Another participant says ``\textit{they 
		are irrelevant for 	the bug}''.
\end{itemize}

\subsection{Threats to Validity}
\label{subsec:limitations}

We  discuss the threats that may affect the validity of the evaluation.

The main threat to the {\em construct validity} is the subjectivity introduced 
in the ideal bug reproduction scenarios given to the study participants. To 
minimize bias, we created them based on the bug reproduction performed by 
third-parties, \ie a group of graduate students. Another threat concerns 
the procedure to assess \approach's usefulness. Our methodology 
only assesses the perceived usefulness of \approach. Investigating how users 
actually benefit from \approach when reporting bugs is subject of future work.

\approach's calibration impacts the  {\em internal validity} of our 
conclusions. As explained in Sec. \ref{subsec:approach-implementation}, we 
used different bug reports (to the ones used in the evaluation) to test and 
find the 
best parameters of the approach. Another threat is subjectivity and 
error-proneness of the human-based evaluation. To mitigate this threat, we 
relied on three evaluators per bug report, and decided upon majority.  

Given a relatively expensive nature of our evaluation, we 
limited it to 24 bug reports and three evaluators for each report, 
which affects the   {\em external validity} of our conclusions. A 
larger evaluation, possibly performed by a diverse (in terms of experience) 
sample of evaluators on additional bug reports, would be desirable.


\section{Related Work}
\label{sec:related-work}

\paragraph{Bug Report Quality Assessment.} 
Zimmermann  \etal~\cite{Zimmermann2010} conducted 
a survey exploring the most useful information in bug reports and 
proposed a supervised approach to predict the overall quality level of a bug 
report (\ie bad, neutral, or good). This approach relies on features, 
such as readability, presence of certain keywords, code snippets, \etc Dit 
\etal~\cite{Dit2008} and Linstead \etal~\cite{Linstead2009} 
measured the semantic coherence in bug report discussions based on textual 
similarity and topic models. Hooimeijer \etal~\cite{Hooimeijer2007} measured 
quality properties of  bug reports (\eg readability) to predict when a report 
would be triaged.  Zanetti \etal~\cite{Zanetti2013} identified valid  bug 
reports, as opposed to duplicate, invalid, or incomplete, by relying on 
reporters' collaboration information. 
To enhance bug reports, Moran \etal focused on 
augmenting S2Rs via screenshots and GUI-component 
images~\cite{Moran2015}. Zhang \etal 
\cite{Zhang2017} enriched new bug reports with textually similar sentences from 
past reports.

\paragraph{Textual Analysis of Bug Reports.} Existing research focused on 
determining the structure of bug reports and its importance in bug triaging and fixing
\cite{3Bettenburg:FSE08,Davies2014,Zimmermann2009,Sasso2016,Sahoo2010, Zimmermann2010,Laukkanen2011}.  
The work by Ko \etal~\cite{Ko2006} on linguistic analysis of bug report 
titles is  complemented by 
 discourse pattern identification in bug descriptions \cite{Chaparro2017-2}. Sureka 
\etal~\cite{Sureka2010} analyzed the part-of-speech and distribution of words 
in titles to find vocabulary patterns for predicting bug severity.  

\paragraph{Test Case Generation and Crash Reproduction from Bug Reports.} 
Fazzini 
\etal 
\cite{Fazzini2018} and Karag{\"o}z \etal \cite{Karagoez2017} proposed 
approaches to generate executable test cases from bug 
reports. Zhao \etal \cite{Zhao2019} proposed a technique to reproduce crashes 
from bug reports.  Different from these approaches, \approach is capable of 
automatically identifying S2Rs in free-form bug report text, and inferring 
 steps missing in the report. \approach is complementary to these techniques, 
 as 
it is aimed at improving the quality of reported S2Rs.  High-quality S2Rs can 
help improve the 
effectiveness of these approaches.

\section{Conclusions and Future Work}
\label{sec:conclusion}

\approach is an approach for the automated quality 
assessment of the steps to reproduce (S2Rs) in bug reports.  \approach 
identifies individual S2Rs in bug reports with high accuracy (98\%), and 
produces a quality report (QR), where for each S2R, provides quality 
annotations (QAs), indicating whether the S2R is well-written, 
ambiguous, or uses unusual vocabulary. The QR includes a list 
of missing S2Rs, automatically inferred by \approach, that are needed for 
reproducing the 
reported bug. External 
evaluators found the QRs easy to understand (they agreed in 87\% 
of the cases), while they rated the accuracy of the QAs. 73\% of the QAs 
were deemed accurate, while \approach reported 58\% of the missing S2Rs (albeit 
with a 31\% precision). The evaluators consider \euler to be potentially useful 
(they agreed in 83\% of the cases) in helping reporters improve their  
bug reports.

Future extrinsic studies will confirm the reported perceived usefulness. Before 
such studies, improvements to \approach's accuracy and to the 
information included in the QRs are planned, based on the evaluators' 
feedback.  Specifically, we plan to improve the quality of 
\approach's QR, 
with: (i) more complete application step sequences; and (ii) additional 
screenshots to help guide reporters. We also plan to tune \approach's 
matching algorithm to account for minor variations between text sequences and 
matches in different parts of the execution model. As discussed in Section  
\ref{subsec:results}, optimizations to the application exploration strategies 
are also planned.

\begin{acks}

This work was partially supported by the NSF grants IIS-1528037 and 
CCF-1815186, 1815336, 1525902, 1848608, and 1526118. 
\end{acks}

\balance
\bibliographystyle{ACM-Reference-Format}
\bibliography{ms}


\begin{thebibliography}{60}


\ifx \showCODEN    \undefined \def \showCODEN     #1{\unskip}     \fi
\ifx \showDOI      \undefined \def \showDOI       #1{#1}\fi
\ifx \showISBNx    \undefined \def \showISBNx     #1{\unskip}     \fi
\ifx \showISBNxiii \undefined \def \showISBNxiii  #1{\unskip}     \fi
\ifx \showISSN     \undefined \def \showISSN      #1{\unskip}     \fi
\ifx \showLCCN     \undefined \def \showLCCN      #1{\unskip}     \fi
\ifx \shownote     \undefined \def \shownote      #1{#1}          \fi
\ifx \showarticletitle \undefined \def \showarticletitle #1{#1}   \fi
\ifx \showURL      \undefined \def \showURL       {\relax}        \fi
\providecommand\bibfield[2]{#2}
\providecommand\bibinfo[2]{#2}
\providecommand\natexlab[1]{#1}
\providecommand\showeprint[2][]{arXiv:#2}

\bibitem[\protect\citeauthoryear{??}{Git}{2016}]%
        {GitHub2016}
 \bibinfo{year}{2016}\natexlab{}.
\newblock \bibinfo{title}{An open letter to GitHub from the maintainers of open
  source projects}.
\newblock
  \bibinfo{howpublished}{\url{https://github.com/dear-github/dear-github}}.
\newblock


\bibitem[\protect\citeauthoryear{??}{gnu}{2017}]%
        {gnucash_bug}
 \bibinfo{year}{2017}\natexlab{}.
\newblock \bibinfo{title}{GnuCash's bug report \#256}.
\newblock \bibinfo{howpublished}{\url{https://tinyurl.com/y3df92g7}}.
\newblock


\bibitem[\protect\citeauthoryear{??}{aar}{2019a}]%
        {aardict}
 \bibinfo{year}{2019}\natexlab{a}.
\newblock \bibinfo{title}{Aard Dictionary}.
\newblock \bibinfo{howpublished}{\url{https://tinyurl.com/mwpxshz}}.
\newblock


\bibitem[\protect\citeauthoryear{??}{aar}{2019b}]%
        {aardic_104}
 \bibinfo{year}{2019}\natexlab{b}.
\newblock \bibinfo{title}{Aard Dictionary's bug report \#104}.
\newblock \bibinfo{howpublished}{\url{https://tinyurl.com/y3xhlky3}}.
\newblock


\bibitem[\protect\citeauthoryear{??}{aar}{2019c}]%
        {aardic_81}
 \bibinfo{year}{2019}\natexlab{c}.
\newblock \bibinfo{title}{Aard Dictionary's bug report \#81}.
\newblock \bibinfo{howpublished}{\url{https://tinyurl.com/y3xvqf3j}}.
\newblock


\bibitem[\protect\citeauthoryear{??}{dro}{2019}]%
        {droidweight}
 \bibinfo{year}{2019}\natexlab{}.
\newblock \bibinfo{title}{Droid Weight}.
\newblock \bibinfo{howpublished}{\url{https://tinyurl.com/lxazk36}}.
\newblock


\bibitem[\protect\citeauthoryear{??}{gnu}{2019a}]%
        {gnucash}
 \bibinfo{year}{2019}\natexlab{a}.
\newblock \bibinfo{title}{GnuCash}.
\newblock \bibinfo{howpublished}{\url{https://tinyurl.com/ku9dqq8}}.
\newblock


\bibitem[\protect\citeauthoryear{??}{gnu}{2019b}]%
        {gnucash_471}
 \bibinfo{year}{2019}\natexlab{b}.
\newblock \bibinfo{title}{GnuCash's bug report \#471}.
\newblock \bibinfo{howpublished}{\url{https://tinyurl.com/y6luonwp}}.
\newblock


\bibitem[\protect\citeauthoryear{??}{gnu}{2019c}]%
        {gnucash_616}
 \bibinfo{year}{2019}\natexlab{c}.
\newblock \bibinfo{title}{GnuCash's bug report \#616}.
\newblock \bibinfo{howpublished}{\url{https://tinyurl.com/y5edsasv}}.
\newblock


\bibitem[\protect\citeauthoryear{??}{gnu}{2019d}]%
        {gnucash_620}
 \bibinfo{year}{2019}\natexlab{d}.
\newblock \bibinfo{title}{GnuCash's bug report \#620}.
\newblock \bibinfo{howpublished}{\url{https://tinyurl.com/y3pw69ac}}.
\newblock


\bibitem[\protect\citeauthoryear{??}{gnu}{2019e}]%
        {gnucash_701}
 \bibinfo{year}{2019}\natexlab{e}.
\newblock \bibinfo{title}{GnuCash's bug report \#701}.
\newblock \bibinfo{howpublished}{\url{https://tinyurl.com/y4e4ny9a}}.
\newblock


\bibitem[\protect\citeauthoryear{??}{mil}{2019a}]%
        {mileage}
 \bibinfo{year}{2019}\natexlab{a}.
\newblock \bibinfo{title}{Mileage}.
\newblock \bibinfo{howpublished}{\url{https://tinyurl.com/cw3uttu}}.
\newblock


\bibitem[\protect\citeauthoryear{??}{mil}{2019b}]%
        {mileage_53}
 \bibinfo{year}{2019}\natexlab{b}.
\newblock \bibinfo{title}{Mileage's bug report \#53}.
\newblock \bibinfo{howpublished}{\url{https://tinyurl.com/y6mo92cm}}.
\newblock


\bibitem[\protect\citeauthoryear{??}{rep}{2019}]%
        {repl_pack}
 \bibinfo{year}{2019}\natexlab{}.
\newblock \bibinfo{title}{Online replication package}.
\newblock
  \bibinfo{howpublished}{\url{https://seers.utdallas.edu/projects/s2r-quality}}.
\newblock


\bibitem[\protect\citeauthoryear{??}{qua}{2019}]%
        {qualtrics}
 \bibinfo{year}{2019}\natexlab{}.
\newblock \bibinfo{title}{Qualtrics online survey system}.
\newblock \bibinfo{howpublished}{\url{https://tinyurl.com/y4fumc6g}}.
\newblock


\bibitem[\protect\citeauthoryear{??}{sch}{2019a}]%
        {schedule}
 \bibinfo{year}{2019}\natexlab{a}.
\newblock \bibinfo{title}{Schedule}.
\newblock \bibinfo{howpublished}{\url{https://tinyurl.com/bsw89ud}}.
\newblock


\bibitem[\protect\citeauthoryear{??}{sch}{2019b}]%
        {schedule_154}
 \bibinfo{year}{2019}\natexlab{b}.
\newblock \bibinfo{title}{Schedule's bug report \#154}.
\newblock \bibinfo{howpublished}{\url{https://tinyurl.com/y3pg92fr}}.
\newblock


\bibitem[\protect\citeauthoryear{??}{sch}{2019c}]%
        {schedule_169}
 \bibinfo{year}{2019}\natexlab{c}.
\newblock \bibinfo{title}{Schedule's bug report \#169}.
\newblock \bibinfo{howpublished}{\url{https://tinyurl.com/y46l44vr}}.
\newblock


\bibitem[\protect\citeauthoryear{??}{ati}{2019a}]%
        {atimetracker}
 \bibinfo{year}{2019}\natexlab{a}.
\newblock \bibinfo{title}{A Time Tracker}.
\newblock \bibinfo{howpublished}{\url{https://tinyurl.com/lt4ztgp}}.
\newblock


\bibitem[\protect\citeauthoryear{??}{ati}{2019b}]%
        {atimetracker_1}
 \bibinfo{year}{2019}\natexlab{b}.
\newblock \bibinfo{title}{A Time Tracker's bug report \#1}.
\newblock \bibinfo{howpublished}{\url{https://tinyurl.com/y4skjrp6}}.
\newblock


\bibitem[\protect\citeauthoryear{??}{ati}{2019c}]%
        {atimetracker_10}
 \bibinfo{year}{2019}\natexlab{c}.
\newblock \bibinfo{title}{A Time Tracker's bug report \#10}.
\newblock \bibinfo{howpublished}{\url{https://tinyurl.com/y4a698hb}}.
\newblock


\bibitem[\protect\citeauthoryear{??}{ati}{2019d}]%
        {atimetracker_35}
 \bibinfo{year}{2019}\natexlab{d}.
\newblock \bibinfo{title}{A Time Tracker's bug report \#35}.
\newblock \bibinfo{howpublished}{\url{https://tinyurl.com/y3tvylgs}}.
\newblock


\bibitem[\protect\citeauthoryear{Baek and Bae}{Baek and Bae}{2016}]%
        {Baek2016}
\bibfield{author}{\bibinfo{person}{Young-Min Baek} {and}
  \bibinfo{person}{Doo-Hwan Bae}.} \bibinfo{year}{2016}\natexlab{}.
\newblock \showarticletitle{Automated Model-based Android GUI Testing Using
  Multi-level GUI Comparison Criteria}. In
  \bibinfo{booktitle}{\emph{Proceedings of the 31st IEEE/ACM International
  Conference on Automated Software Engineering}}
  \emph{(\bibinfo{series}{ASE'16})}. \bibinfo{pages}{238--249}.
\newblock
\showISBNx{978-1-4503-3845-5}


\bibitem[\protect\citeauthoryear{Bettenburg, Just, Schr\"{o}ter, Weiss,
  Premraj, and Zimmermann}{Bettenburg et~al\mbox{.}}{2008}]%
        {3Bettenburg:FSE08}
\bibfield{author}{\bibinfo{person}{Nicolas Bettenburg}, \bibinfo{person}{Sascha
  Just}, \bibinfo{person}{Adrian Schr\"{o}ter}, \bibinfo{person}{Cathrin
  Weiss}, \bibinfo{person}{Rahul Premraj}, {and} \bibinfo{person}{Thomas
  Zimmermann}.} \bibinfo{year}{2008}\natexlab{}.
\newblock \showarticletitle{What Makes a Good Bug Report?}. In
  \bibinfo{booktitle}{\emph{Proceedings of the 16th International Symposium on
  the Foundations of Software Engineering}} \emph{(\bibinfo{series}{FSE'08})}.
  \bibinfo{pages}{308--318}.
\newblock
\showISBNx{978-1-59593-995-1}


\bibitem[\protect\citeauthoryear{Bojanowski, Grave, Joulin, and
  Mikolov}{Bojanowski et~al\mbox{.}}{2017}]%
        {fasttext}
\bibfield{author}{\bibinfo{person}{Piotr Bojanowski}, \bibinfo{person}{Edouard
  Grave}, \bibinfo{person}{Armand Joulin}, {and} \bibinfo{person}{Tomas
  Mikolov}.} \bibinfo{year}{2017}\natexlab{}.
\newblock \showarticletitle{Enriching Word Vectors with Subword Information}.
\newblock \bibinfo{journal}{\emph{Transactions of the Association for
  Computational Linguistics}}  \bibinfo{volume}{5} (\bibinfo{year}{2017}),
  \bibinfo{pages}{135--146}.
\newblock


\bibitem[\protect\citeauthoryear{Breu, Premraj, Sillito, and Zimmermann}{Breu
  et~al\mbox{.}}{2010}]%
        {Breu2010}
\bibfield{author}{\bibinfo{person}{Silvia Breu}, \bibinfo{person}{Rahul
  Premraj}, \bibinfo{person}{Jonathan Sillito}, {and} \bibinfo{person}{Thomas
  Zimmermann}.} \bibinfo{year}{2010}\natexlab{}.
\newblock \showarticletitle{Information {Needs} in {Bug} {Reports}: {Improving}
  {Cooperation} {Between} {Developers} and {Users}}. In
  \bibinfo{booktitle}{\emph{Proceedings of the {Conference} on {Computer}
  {Supported} {Cooperative} {Work} (CSCW'10)}}. \bibinfo{pages}{301--310}.
\newblock


\bibitem[\protect\citeauthoryear{Chaparro, Lu, Zampetti, Moreno, Di~Penta,
  Marcus, Bavota, and Ng}{Chaparro et~al\mbox{.}}{2017}]%
        {Chaparro2017-2}
\bibfield{author}{\bibinfo{person}{Oscar Chaparro}, \bibinfo{person}{Jing Lu},
  \bibinfo{person}{Fiorella Zampetti}, \bibinfo{person}{Laura Moreno},
  \bibinfo{person}{Massimiliano Di~Penta}, \bibinfo{person}{Andrian Marcus},
  \bibinfo{person}{Gabriele Bavota}, {and} \bibinfo{person}{Vincent Ng}.}
  \bibinfo{year}{2017}\natexlab{}.
\newblock \showarticletitle{Detecting Missing Information in Bug Descriptions}.
  In \bibinfo{booktitle}{\emph{Proceedings of the 11th Joint Meeting on the
  Foundations of Software Engineering (ESEC/FSE'17)}}.
  \bibinfo{pages}{396--407}.
\newblock


\bibitem[\protect\citeauthoryear{Conneau, Kruszewski, Lample, Barrault, and
  Baroni}{Conneau et~al\mbox{.}}{2018}]%
        {sentembed}
\bibfield{author}{\bibinfo{person}{Alexis Conneau},
  \bibinfo{person}{Germ{\'{a}}n Kruszewski}, \bibinfo{person}{Guillaume
  Lample}, \bibinfo{person}{Lo{\"{\i}}c Barrault}, {and} \bibinfo{person}{Marco
  Baroni}.} \bibinfo{year}{2018}\natexlab{}.
\newblock \showarticletitle{What you can cram into a single vector: Probing
  sentence embeddings for linguistic properties}.
\newblock \bibinfo{journal}{\emph{CoRR}}  \bibinfo{volume}{abs/1805.01070}
  (\bibinfo{year}{2018}).
\newblock


\bibitem[\protect\citeauthoryear{Davies and Roper}{Davies and Roper}{2014}]%
        {Davies2014}
\bibfield{author}{\bibinfo{person}{Steven Davies} {and} \bibinfo{person}{Marc
  Roper}.} \bibinfo{year}{2014}\natexlab{}.
\newblock \showarticletitle{What's in a bug report?}. In
  \bibinfo{booktitle}{\emph{Proceedings of the 8th International Symposium on
  Empirical Software Engineering and Measurement}}
  \emph{(\bibinfo{series}{ESEM'14})}. \bibinfo{pages}{26:1--26:10}.
\newblock


\bibitem[\protect\citeauthoryear{Dit, Poshyvanyk, and Marcus}{Dit
  et~al\mbox{.}}{2008}]%
        {Dit2008}
\bibfield{author}{\bibinfo{person}{Bogdan Dit}, \bibinfo{person}{Denys
  Poshyvanyk}, {and} \bibinfo{person}{Andrian Marcus}.}
  \bibinfo{year}{2008}\natexlab{}.
\newblock \showarticletitle{Measuring the semantic similarity of comments in
  bug reports}. In \bibinfo{booktitle}{\emph{Proceedings of the 1st
  International Workshop on Semantic Technologies in System Maintenance
  (STSM'08)}}. \bibinfo{pages}{265--280}.
\newblock


\bibitem[\protect\citeauthoryear{Erfani~Joorabchi, Mirzaaghaei, and
  Mesbah}{Erfani~Joorabchi et~al\mbox{.}}{2014}]%
        {ErfaniJoorabchi2014}
\bibfield{author}{\bibinfo{person}{Mona Erfani~Joorabchi},
  \bibinfo{person}{Mehdi Mirzaaghaei}, {and} \bibinfo{person}{Ali Mesbah}.}
  \bibinfo{year}{2014}\natexlab{}.
\newblock \showarticletitle{Works for {Me}! {Characterizing} {Non}-reproducible
  {Bug} {Reports}}. In \bibinfo{booktitle}{\emph{Proceedings of the {Working}
  {Conference} on {Mining} {Software} {Repositories} (MSR'14)}}.
  \bibinfo{pages}{62--71}.
\newblock


\bibitem[\protect\citeauthoryear{Fazzini, Prammer, d'Amorim, and Orso}{Fazzini
  et~al\mbox{.}}{2018}]%
        {Fazzini2018}
\bibfield{author}{\bibinfo{person}{Mattia Fazzini}, \bibinfo{person}{Martin
  Prammer}, \bibinfo{person}{Marcelo d'Amorim}, {and}
  \bibinfo{person}{Alessandro Orso}.} \bibinfo{year}{2018}\natexlab{}.
\newblock \showarticletitle{Automatically translating bug reports into test
  cases for mobile apps}. In \bibinfo{booktitle}{\emph{Proceedings of the 27th
  International Symposium on Software Testing and Analysis (ISSTA'18)}}.
  \bibinfo{pages}{141--152}.
\newblock


\bibitem[\protect\citeauthoryear{Guo, Zimmermann, Nagappan, and Murphy}{Guo
  et~al\mbox{.}}{2010}]%
        {Guo2010}
\bibfield{author}{\bibinfo{person}{Philip~J. Guo}, \bibinfo{person}{Thomas
  Zimmermann}, \bibinfo{person}{Nachiappan Nagappan}, {and}
  \bibinfo{person}{Brendan Murphy}.} \bibinfo{year}{2010}\natexlab{}.
\newblock \showarticletitle{Characterizing and predicting which bugs get fixed:
  an empirical study of Microsoft Windows}. In
  \bibinfo{booktitle}{\emph{Proceedings of the 32nd International Conference on
  Software Engineering}} \emph{(\bibinfo{series}{ICSE'10})},
  Vol.~\bibinfo{volume}{1}. \bibinfo{pages}{495--504}.
\newblock


\bibitem[\protect\citeauthoryear{Hooimeijer and Weimer}{Hooimeijer and
  Weimer}{2007}]%
        {Hooimeijer2007}
\bibfield{author}{\bibinfo{person}{Pieter Hooimeijer} {and}
  \bibinfo{person}{Westley Weimer}.} \bibinfo{year}{2007}\natexlab{}.
\newblock \showarticletitle{Modeling {Bug} {Report} {Quality}}. In
  \bibinfo{booktitle}{\emph{Proceedings of the 22nd International Conference on
  Automated Software Engineering (ASE'07)}}. \bibinfo{pages}{34--43}.
\newblock


\bibitem[\protect\citeauthoryear{Huang, Xu, and Yu}{Huang
  et~al\mbox{.}}{2015}]%
        {Huang2016LSTM}
\bibfield{author}{\bibinfo{person}{Zhiheng Huang}, \bibinfo{person}{Wei Xu},
  {and} \bibinfo{person}{Kai Yu}.} \bibinfo{year}{2015}\natexlab{}.
\newblock \showarticletitle{Bidirectional {LSTM-CRF} Models for Sequence
  Tagging}.
\newblock \bibinfo{journal}{\emph{CoRR}}  \bibinfo{volume}{abs/1508.01991}
  (\bibinfo{year}{2015}).
\newblock


\bibitem[\protect\citeauthoryear{Karag{\"o}z and S{\"o}zer}{Karag{\"o}z and
  S{\"o}zer}{2017}]%
        {Karagoez2017}
\bibfield{author}{\bibinfo{person}{G{\"u}n Karag{\"o}z} {and}
  \bibinfo{person}{Hasan S{\"o}zer}.} \bibinfo{year}{2017}\natexlab{}.
\newblock \showarticletitle{Reproducing failures based on semiformal failure
  scenario descriptions}.
\newblock \bibinfo{journal}{\emph{Software Quality Journal}}
  \bibinfo{volume}{25}, \bibinfo{number}{1} (\bibinfo{year}{2017}),
  \bibinfo{pages}{111--129}.
\newblock


\bibitem[\protect\citeauthoryear{Ko, Myers, and Chau}{Ko et~al\mbox{.}}{2006}]%
        {Ko2006}
\bibfield{author}{\bibinfo{person}{Andrew~J. Ko}, \bibinfo{person}{Brad~A
  Myers}, {and} \bibinfo{person}{Duen~Horng Chau}.}
  \bibinfo{year}{2006}\natexlab{}.
\newblock \showarticletitle{A {Linguistic} {Analysis} of {How} {People}
  {Describe} {Software} {Problems}}. In \bibinfo{booktitle}{\emph{Proceedings
  of the Symposium on Visual Languages and Human-Centric Computing
  (VL/HCC'06)}}. \bibinfo{pages}{127--134}.
\newblock


\bibitem[\protect\citeauthoryear{Lample, Ballesteros, Subramanian, Kawakami,
  and Dyer}{Lample et~al\mbox{.}}{2016}]%
        {Lample2016neural}
\bibfield{author}{\bibinfo{person}{Guillaume Lample}, \bibinfo{person}{Miguel
  Ballesteros}, \bibinfo{person}{Sandeep Subramanian}, \bibinfo{person}{Kazuya
  Kawakami}, {and} \bibinfo{person}{Chris Dyer}.}
  \bibinfo{year}{2016}\natexlab{}.
\newblock \showarticletitle{Neural Architectures for Named Entity Recognition}.
  In \bibinfo{booktitle}{\emph{Proceedings of North American Chapter of the
  Association for Computational Linguistics: Human Language Technologies
  (NAACL-HLT'16)}}. \bibinfo{pages}{260--270}.
\newblock


\bibitem[\protect\citeauthoryear{Laukkanen and M\"{a}ntyl\"{a}}{Laukkanen and
  M\"{a}ntyl\"{a}}{2011}]%
        {Laukkanen2011}
\bibfield{author}{\bibinfo{person}{Eero~I. Laukkanen} {and}
  \bibinfo{person}{Mika~V. M\"{a}ntyl\"{a}}.} \bibinfo{year}{2011}\natexlab{}.
\newblock \showarticletitle{Survey Reproduction of Defect Reporting in
  Industrial Software Development}. In \bibinfo{booktitle}{\emph{Proceedings of
  the International Symposium on Empirical Software Engineering and
  Measurement}} \emph{(\bibinfo{series}{ESEM'11})}. \bibinfo{pages}{197--206}.
\newblock


\bibitem[\protect\citeauthoryear{Linstead and Baldi}{Linstead and
  Baldi}{2009}]%
        {Linstead2009}
\bibfield{author}{\bibinfo{person}{Erik Linstead} {and} \bibinfo{person}{Pierre
  Baldi}.} \bibinfo{year}{2009}\natexlab{}.
\newblock \showarticletitle{Mining the coherence of GNOME bug reports with
  statistical topic models}. In \bibinfo{booktitle}{\emph{Proceedings of the
  6th International Working Conference on Mining Software Repositories
  (MSR'09)}}. \bibinfo{pages}{99--102}.
\newblock


\bibitem[\protect\citeauthoryear{Ma and Hovy}{Ma and Hovy}{2016}]%
        {ma2016end}
\bibfield{author}{\bibinfo{person}{Xuezhe Ma} {and} \bibinfo{person}{Eduard
  Hovy}.} \bibinfo{year}{2016}\natexlab{}.
\newblock \showarticletitle{End-to-end Sequence Labeling via Bi-directional
  LSTM-CNNs-CRF}. In \bibinfo{booktitle}{\emph{Proceedings of the 54th Annual
  Meeting of the Association for Computational Linguistics}}
  \emph{(\bibinfo{series}{ACL'16})}, Vol.~\bibinfo{volume}{1}.
  \bibinfo{pages}{1064--1074}.
\newblock


\bibitem[\protect\citeauthoryear{Manning, Surdeanu, Bauer, Finkel, Bethard, and
  McClosky}{Manning et~al\mbox{.}}{2014}]%
        {Manning2014}
\bibfield{author}{\bibinfo{person}{Christopher~D Manning},
  \bibinfo{person}{Mihai Surdeanu}, \bibinfo{person}{John Bauer},
  \bibinfo{person}{Jenny~Rose Finkel}, \bibinfo{person}{Steven Bethard}, {and}
  \bibinfo{person}{David McClosky}.} \bibinfo{year}{2014}\natexlab{}.
\newblock \showarticletitle{The Stanford CoreNLP Natural Language Processing
  Toolkit}. In \bibinfo{booktitle}{\emph{Proceedings of the 52nd Annual Meeting
  of the Association for Computational Linguistics}}
  \emph{(\bibinfo{series}{ACL'14})}. \bibinfo{pages}{55--60}.
\newblock


\bibitem[\protect\citeauthoryear{Moran, Linares-V\'{a}quez,
  Bernal-C\'{a}rdenas, Vendome, and Poshyvanyk}{Moran et~al\mbox{.}}{2016}]%
        {Moran2016}
\bibfield{author}{\bibinfo{person}{Kevin Moran}, \bibinfo{person}{Mario
  Linares-V\'{a}quez}, \bibinfo{person}{Carlos Bernal-C\'{a}rdenas},
  \bibinfo{person}{Christopher Vendome}, {and} \bibinfo{person}{Denys
  Poshyvanyk}.} \bibinfo{year}{2016}\natexlab{}.
\newblock \showarticletitle{Automatically {Discovering}, {Reporting} and
  {Reproducing} {Android} {Application} {Crashes}}. In
  \bibinfo{booktitle}{\emph{Proceedings of the {International} {Conference} on
  {Software} {Testing}, {Verification} and {Validation} (ICST'16)}}.
  \bibinfo{pages}{33--44}.
\newblock


\bibitem[\protect\citeauthoryear{Moran, Linares-V\'{a}squez,
  Bernal-C\'{a}rdenas, and Poshyvanyk}{Moran et~al\mbox{.}}{2015}]%
        {Moran2015}
\bibfield{author}{\bibinfo{person}{Kevin Moran}, \bibinfo{person}{Mario
  Linares-V\'{a}squez}, \bibinfo{person}{Carlos Bernal-C\'{a}rdenas}, {and}
  \bibinfo{person}{Denys Poshyvanyk}.} \bibinfo{year}{2015}\natexlab{}.
\newblock \showarticletitle{Auto-completing {Bug} {Reports} for {Android}
  {Applications}}. In \bibinfo{booktitle}{\emph{Proceedings of the {Joint}
  {Meeting} on {Foundations} of {Software} {Engineering} (FSE'15)}}.
  \bibinfo{pages}{673--686}.
\newblock


\bibitem[\protect\citeauthoryear{{Moran}, {Linares-Vasquez}, {Bernal-Cardenas},
  {Vendome}, and {Poshyvanyk}}{{Moran} et~al\mbox{.}}{2017}]%
        {Moran:2017}
\bibfield{author}{\bibinfo{person}{Kevin {Moran}}, \bibinfo{person}{Mario
  {Linares-Vasquez}}, \bibinfo{person}{Carlos {Bernal-Cardenas}},
  \bibinfo{person}{Cristopher {Vendome}}, {and} \bibinfo{person}{Denys
  {Poshyvanyk}}.} \bibinfo{year}{2017}\natexlab{}.
\newblock \showarticletitle{CrashScope: A Practical Tool for Automated Testing
  of Android Applications}. In \bibinfo{booktitle}{\emph{Proceedings of the
  IEEE/ACM 39th International Conference on Software Engineering (ICSE'17)}}.
  \bibinfo{pages}{15--18}.
\newblock


\bibitem[\protect\citeauthoryear{Oppenheim}{Oppenheim}{1992}]%
        {Oppenheim:1992}
\bibfield{author}{\bibinfo{person}{Abraham~Naftali Oppenheim}.}
  \bibinfo{year}{1992}\natexlab{}.
\newblock \bibinfo{booktitle}{\emph{Questionnaire Design, Interviewing and
  Attitude Measurement}}.
\newblock \bibinfo{publisher}{Pinter Publishers}.
\newblock


\bibitem[\protect\citeauthoryear{Ramshaw and Marcus}{Ramshaw and
  Marcus}{1999}]%
        {Ramshaw1999NLP}
\bibfield{author}{\bibinfo{person}{Lance~A Ramshaw} {and}
  \bibinfo{person}{Mitchell~P Marcus}.} \bibinfo{year}{1999}\natexlab{}.
\newblock \showarticletitle{Text chunking using transformation-based learning}.
\newblock In \bibinfo{booktitle}{\emph{Natural language processing using very
  large corpora}}. \bibinfo{pages}{157--176}.
\newblock


\bibitem[\protect\citeauthoryear{Sahoo, Criswell, and Adve}{Sahoo
  et~al\mbox{.}}{2010}]%
        {Sahoo2010}
\bibfield{author}{\bibinfo{person}{Swarup~Kumar Sahoo}, \bibinfo{person}{John
  Criswell}, {and} \bibinfo{person}{Vikram Adve}.}
  \bibinfo{year}{2010}\natexlab{}.
\newblock \showarticletitle{An empirical study of reported bugs in server
  software with implications for automated bug diagnosis}. In
  \bibinfo{booktitle}{\emph{Proceedings of the International Conference on
  Software Engineering (ICSE'10)}}. \bibinfo{pages}{485--494}.
\newblock


\bibitem[\protect\citeauthoryear{Sasso, Mocci, and Lanza}{Sasso
  et~al\mbox{.}}{2016}]%
        {Sasso2016}
\bibfield{author}{\bibinfo{person}{Tommaso~Dal Sasso}, \bibinfo{person}{Andrea
  Mocci}, {and} \bibinfo{person}{Michele Lanza}.}
  \bibinfo{year}{2016}\natexlab{}.
\newblock \showarticletitle{What {Makes} a {Satisficing} {Bug} {Report}?}. In
  \bibinfo{booktitle}{\emph{Proceedings of the {International} {Conference} on
  {Software} {Quality}, {Reliability} and {Security} (QRS'16)}}.
  \bibinfo{pages}{164--174}.
\newblock


\bibitem[\protect\citeauthoryear{Spencer}{Spencer}{2009}]%
        {cardSorting}
\bibfield{author}{\bibinfo{person}{Donna Spencer}.}
  \bibinfo{year}{2009}\natexlab{}.
\newblock \bibinfo{booktitle}{\emph{Card sorting: Designing usable
  categories}}.
\newblock \bibinfo{publisher}{Rosenfeld Media}.
\newblock


\bibitem[\protect\citeauthoryear{Sureka and Jalote}{Sureka and Jalote}{2010}]%
        {Sureka2010}
\bibfield{author}{\bibinfo{person}{Ashish Sureka} {and} \bibinfo{person}{Pankaj
  Jalote}.} \bibinfo{year}{2010}\natexlab{}.
\newblock \showarticletitle{Detecting {Duplicate} {Bug} {Report} {Using}
  {Character} {N}-{Gram}-{Based} {Features}}. In
  \bibinfo{booktitle}{\emph{Proceedings of the {Asia} {Pacific} {Software}
  {Engineering} {Conference} (APSEC'10)}}. \bibinfo{pages}{366--374}.
\newblock


\bibitem[\protect\citeauthoryear{Yang, Liang, and Zhang}{Yang
  et~al\mbox{.}}{2018}]%
        {neuralseq}
\bibfield{author}{\bibinfo{person}{Jie Yang}, \bibinfo{person}{Shuailong
  Liang}, {and} \bibinfo{person}{Yue Zhang}.} \bibinfo{year}{2018}\natexlab{}.
\newblock \showarticletitle{Design Challenges and Misconceptions in Neural
  Sequence Labeling}. In \bibinfo{booktitle}{\emph{Proceedings of the 27th
  International Conference on Computational Linguistics (COLING'18)}}.
  \bibinfo{pages}{3879--3889}.
\newblock


\bibitem[\protect\citeauthoryear{Yang and Zhang}{Yang and Zhang}{2018}]%
        {ncrf}
\bibfield{author}{\bibinfo{person}{Jie Yang} {and} \bibinfo{person}{Yue
  Zhang}.} \bibinfo{year}{2018}\natexlab{}.
\newblock \showarticletitle{NCRF++: An Open-source Neural Sequence Labeling
  Toolkit}. In \bibinfo{booktitle}{\emph{Proceedings of the 56th Annual Meeting
  of the Association for Computational Linguistics (ACL'18)}}.
\newblock


\bibitem[\protect\citeauthoryear{Zaeem, Prasad, and Khurshid}{Zaeem
  et~al\mbox{.}}{2014}]%
        {Zaeem2014}
\bibfield{author}{\bibinfo{person}{Razieh~Nokhbeh Zaeem},
  \bibinfo{person}{Mukul~R. Prasad}, {and} \bibinfo{person}{Sarfraz Khurshid}.}
  \bibinfo{year}{2014}\natexlab{}.
\newblock \showarticletitle{Automated Generation of Oracles for Testing
  User-Interaction Features of Mobile Apps}. In
  \bibinfo{booktitle}{\emph{Proceedings of the 7th International Conference on
  Software Testing, Verification and Validation (ICST'14)}}.
  \bibinfo{pages}{183--192}.
\newblock


\bibitem[\protect\citeauthoryear{Zanetti, Scholtes, Tessone, and
  Schweitzer}{Zanetti et~al\mbox{.}}{2013}]%
        {Zanetti2013}
\bibfield{author}{\bibinfo{person}{Marcelo~Serrano Zanetti},
  \bibinfo{person}{Ingo Scholtes}, \bibinfo{person}{Claudio~Juan Tessone},
  {and} \bibinfo{person}{Frank Schweitzer}.} \bibinfo{year}{2013}\natexlab{}.
\newblock \showarticletitle{Categorizing Bugs with Social Networks: A Case
  Study on Four Open Source Software Communities}. In
  \bibinfo{booktitle}{\emph{Proceedings of the International Conference on
  Software Engineering (ICSE'13)}}. \bibinfo{pages}{1032--1041}.
\newblock


\bibitem[\protect\citeauthoryear{Zhang, Chen, Jiang, Luo, and Xia}{Zhang
  et~al\mbox{.}}{2017}]%
        {Zhang2017}
\bibfield{author}{\bibinfo{person}{Tao Zhang}, \bibinfo{person}{Jiachi Chen},
  \bibinfo{person}{He Jiang}, \bibinfo{person}{Xiapu Luo}, {and}
  \bibinfo{person}{Xin Xia}.} \bibinfo{year}{2017}\natexlab{}.
\newblock \showarticletitle{Bug Report Enrichment with Application of Automated
  Fixer Recommendation}. In \bibinfo{booktitle}{\emph{Proceedings of the 25th
  International Conference on Program Comprehension (ICPC'17)}}.
  \bibinfo{pages}{230--240}.
\newblock


\bibitem[\protect\citeauthoryear{Zhao, Yu, Su, Liu, Zheng, Zhang, and
  Halfond}{Zhao et~al\mbox{.}}{2019}]%
        {Zhao2019}
\bibfield{author}{\bibinfo{person}{Yu Zhao}, \bibinfo{person}{Tingting Yu},
  \bibinfo{person}{Ting Su}, \bibinfo{person}{Yang Liu}, \bibinfo{person}{Wei
  Zheng}, \bibinfo{person}{Jingzhi Zhang}, {and} \bibinfo{person}{William~G.J.
  Halfond}.} \bibinfo{year}{2019}\natexlab{}.
\newblock \showarticletitle{ReCDroid: Automatically Reproducing Android
  Application Crashes from Bug Reports}. In
  \bibinfo{booktitle}{\emph{Proceedings of the 41st ACM/IEEE International
  Conference on Software Engineering (ICSE'19)}}. \bibinfo{pages}{128--139}.
\newblock


\bibitem[\protect\citeauthoryear{Zimmermann, Nagappan, Guo, and
  Murphy}{Zimmermann et~al\mbox{.}}{2012}]%
        {Zimmermann2012}
\bibfield{author}{\bibinfo{person}{Thomas Zimmermann},
  \bibinfo{person}{Nachiappan Nagappan}, \bibinfo{person}{Philip~J. Guo}, {and}
  \bibinfo{person}{Brendan Murphy}.} \bibinfo{year}{2012}\natexlab{}.
\newblock \showarticletitle{Characterizing and predicting which bugs get
  reopened}. In \bibinfo{booktitle}{\emph{Proceedings of the International
  Conference on Software Engineering (ICSE'12)}}. \bibinfo{pages}{1074--1083}.
\newblock


\bibitem[\protect\citeauthoryear{Zimmermann, Premraj, Bettenburg, Just,
  Schr\"{o}ter, and Weiss}{Zimmermann et~al\mbox{.}}{2010}]%
        {Zimmermann2010}
\bibfield{author}{\bibinfo{person}{Thomas Zimmermann}, \bibinfo{person}{Rahul
  Premraj}, \bibinfo{person}{Nicolas Bettenburg}, \bibinfo{person}{Sascha
  Just}, \bibinfo{person}{Adrian Schr\"{o}ter}, {and} \bibinfo{person}{Cathrin
  Weiss}.} \bibinfo{year}{2010}\natexlab{}.
\newblock \showarticletitle{What {Makes} a {Good} {Bug} {Report}?}
\newblock \bibinfo{journal}{\emph{IEEE Transactions on Software Engineering}}
  \bibinfo{volume}{36}, \bibinfo{number}{5} (\bibinfo{year}{2010}),
  \bibinfo{pages}{618--643}.
\newblock
\showISSN{0098-5589}


\bibitem[\protect\citeauthoryear{Zimmermann, Premraj, Sillito, and
  Breu}{Zimmermann et~al\mbox{.}}{2009}]%
        {Zimmermann2009}
\bibfield{author}{\bibinfo{person}{Thomas Zimmermann}, \bibinfo{person}{Rahul
  Premraj}, \bibinfo{person}{Jonathan Sillito}, {and} \bibinfo{person}{Silvia
  Breu}.} \bibinfo{year}{2009}\natexlab{}.
\newblock \showarticletitle{Improving bug tracking systems}. In
  \bibinfo{booktitle}{\emph{Proceedings of the 31st International Conference on
  Software Engineering (ICSE'09)}}. \bibinfo{pages}{247--250}.
\newblock


\end{thebibliography}

\end{document}